\documentclass[mrm]{mrm}
\setboolean{bDraft}{false}

\setboolean{bComments}{true}

\usepackage{mathtools}
\usepackage{gensymb}

\makeatletter
\newcommand\newsubcommand[3]{\newcommand#1{#2\sc@sub{#3}}}
\def\sc@sub#1{\def\sc@thesub{#1}\@ifnextchar_{\sc@mergesubs}{_{\sc@thesub}}}
\def\sc@mergesubs_#1{_{\sc@thesub#1}}
\makeatother

\makeatletter
\newcommand\newsupcommand[3]{\newcommand#1{#2\sc@sup{#3}}}
\def\sc@sup#1{\def\sc@thesup{#1}\@ifnextchar^{\sc@mergesups}{^{\sc@thesup}}}
\def\sc@mergesups^#1{^{\sc@thesup#1}}
\makeatother

\newcommand{\argmin}{\operatornamewithlimits{arg\,min}}

\newcommand{\norm}[2][]{\left\|{#2}\right\|_{{#1}}}

\newcommand{\adjoint}{*}

\newcommand{\A}{A}
\newcommand{\AT}{\A^\adjoint}

\newcommand{\reu}{\text{re}}
\newcommand{\imu}{\text{im}}

\newcommand{\FT}{\mathcal{F}}

\newcommand{\IFT}{\mathcal{F}^{-1}}

\newcommand{\x}{x}

\newsubcommand{\xgt}{\x}{\text{ref}}
\newsubcommand{\xre}{\x}{\reu}
\newsubcommand{\xim}{\x}{\imu}
\newsubcommand{\xmem}{\x}{\text{mem}}
\newsupcommand{\xopt}{\x}{*}

\newcommand{\y}{y}
\newsupcommand{\yopt}{\y}{*}

\newcommand{\xdual}{p}
\newsubcommand{\xdualmem}{\xdual}{\text{mem}}

\renewcommand{\t}{t}
\newcommand{\tp}{{t+1}}

\newcommand{\tphalf}{{t+\frac{1}{2}}}

\newcommand{\T}{T}

\newcommand{\data}[1]{\mathcal{D}\left[#1 \right]}

\newsupcommand{\actprime}{\phi}{\prime}

\newcommand{\redequal}{\!=\!}

\def\sota{state-of-the-art\xspace}
\newcommand{\eg}{e.g.,\xspace}
\newcommand{\etal}{et al.\xspace}
\newcommand{\ie}{i.e.,\xspace}

\newcommand{\kspace}{$k$-space\xspace}

\newcommand{\fastmri}{fastMRI\xspace}
\newcommand{\holykspace}{\textit{holykspace}\xspace}

\newcommand{\dnmse}{$\mathcal{D}$-NMSE\xspace}

\newcommand{\unet}{U-net\xspace}
\newcommand{\sn}[2]{SN$_{\text{#1}}^{#2}$}
\newcommand{\pcn}[2]{PCN$_{\text{#1}}^{#2}$}
\newcommand{\sigmanet}{$\Sigma$-net\xspace}
\newcommand{\sigmanetarch}{$\Sigma$-net$^8$\xspace}

\usepackage[smaller]{glossaries}

\setglossarysection{paragraph}

\makeglossaries
\newacronym{acl}{{ACL}}{Auto-Calibration Line}
\newacronym{ai}{{AI}}{Artificial Intelligence}
\newacronym{ann}{{ANN}}{Artificial Neural Network}
\newacronym{aloha}{{ALOHA}}{Annihilating Filter-Based Low-Rank Hankel Structured Matrix Completion Approach}
\newacronym{automap}{{AUTOMAP}}{Automated Transform by Manifold Approximation}
\newacronym{bp}{{BP}}{Backprojection}
\newacronym{admm}{{ADMM}}{Alternating Direction Method of Multipliers}
\newacronym{cnn}{{CNN}}{Convolutional Neural Network}
\newacronym{cyclegan}{{cycleGAN}}{Cycle-Consistent Adversarial Network}
\newacronym{cg}{{CG}}{Conjugate Gradient}
\newacronym{cs}{{CS}}{Compressed Sensing}
\newacronym{ct}{{CT}}{Computed Tomography}
\newacronym{cpu}{{CPU}}{Central Processing Unit}
\newacronym{dicom}{{DICOM}}{Digital Imaging and Communications in Medicine}
\newacronym{dc}{{DC}}{Data Consistency}
\newacronym{dof}{{DOF}}{Degrees of Freedom}
\newacronym{dce}{{DCE}}{Dynamic Contrast Enhanced}
\newacronym{ddr}{{DDR}}{Digitally Reconstructed Radiograph}
\newacronym{dsc}{{DSC}}{Dice Similarity Coefficient}
\newacronym{dun}{{DUN}}{Down-Up Network}
\newacronym{dub}{{DUB}}{Down-Up Block}
\newacronym{dti}{{DTI}}{Diffusion Tensor Imaging}
\newacronym{ecg}{{ECG}}{Electrocardiography}
\newacronym{em}{{EM}}{Expectation Maximization}
\newacronym{epi}{{EPI}}{Echo Planar Imaging}
\newacronym{fe}{{FE}}{Frequency Encoding}
\newacronym{ft}{{FT}}{Fourier Transform}
\newacronym{fft}{{FFT}}{Fast Fourier Transform}
\newacronym{fista}{{FISTA}}{Fast Iterative Shrinkage and Thresholding Algorithm}
\newacronym{fbp}{{FBP}}{Filtered Back-Projection}
\newacronym{fid}{{FID}}{Free Induction Decay}
\newacronym{fsim}{{FSIM}}{Feature Similarity Index}
\newacronym{foe}{{FoE}}{Fields of Experts}
\newacronym{fov}{{FoV}}{Field of View}
\newacronym{fs}{{FS}}{Fat Saturation}
\newacronym{gac}{{GAC}}{Geodesic Active Contours}
\newacronym{gan}{{GAN}}{Generative Adversarial Network}
\newacronym{gd}{{GD}}{Gradient Descent}
\newacronym{gmm}{{GMM}}{Gaussian Mixture Model}
\newacronym{gpu}{{GPU}}{Graphics Processing Unit}
\newacronym{gre}{{GRE}}{Gradient Echo}
\newacronym{grappa}{{GRAPPA}}{Generalized Autocalibrating Partially Parallel Acquisitions}
\newacronym{hu}{{HU}}{Hounsfield Units}
\newacronym{ic}{{IC}}{Infimal Convolution}
\newacronym{ista}{{ISTA}}{Iterative Shrinkage and Thresholding Algorithm}
\newacronym{iipg}{{IIPG}}{Inertial Incremental Proximal Gradient}
\newacronym{ipalm}{{IPALM}}{Inertial Proximal Alternating Linearized Minimization}
\newacronym{ictgv}{{ICTGV}}{Infimal-Convolution-Total-Generalized-Variation}
\newacronym{ksae}{{KSAE}}{K-sparse Autoencoder}
\newacronym{lsc}{{l.s.c.}}{lower-semicontinuous}
\newacronym{lpluss}{{L+S}}{Low-Rank plus Sparse}
\newacronym{lista}{{LISTA}}{Learned Iterative Shrinkage and Thresholding Algorithm}
\newacronym{lbfgs}{{L-BFGS}}{Limited-Memory Broyden-Fletcher-Goldfarb-Shanno}
\newacronym{lsgan}{{LSGAN}}{Least Squares Generative Adversarial Networks}
\newacronym{lvot}{{LVOT}}{Left Ventricular Outflow Tract}
\newacronym{map}{{MAP}}{Maximum-A-Posteriori}
\newacronym{md}{{MD}}{Medical Doctor}
\newacronym{mar}{{MAR}}{Metal Artifact Correction}
\newacronym{mlp}{{MLP}}{Multi Layer Perceptron}
\newacronym{mr}{{MR}}{Magnetic Resonance}
\newacronym{mri}{{MRI}}{Magnetic Resonance Imaging}
\newacronym{mae}{{MAE}}{Mean Absolute Error}
\newacronym{mse}{{MSE}}{Mean Squared Error}
\newacronym{msssim}{{MS-SSIM}}{Multi-Scale Structural Similarity Index}
\newacronym{ncc}{{NCC}}{Normalized Cross Correlation}
\newacronym{nlm}{{NLM}}{Non-Local Means}
\newacronym{nn}{{NN}}{Neural Network}
\newacronym{nufft}{{NUFFT}}{Non-Uniform Fast Fourier Transform}
\newacronym{nrmse}{{NRMSE}}{Normalized Root Mean Squared Error}
\newacronym{nmse}{{NMSE}}{Normalized Mean Squared Error}
\newacronym{icp}{{ICP}}{Iterative Closest Point}
\newacronym{omp}{{OMP}}{Orthogonal Matching Pursuit}
\newacronym{pat}{{PAT}}{Photoacoustic Tomography}
\newacronym{pca}{{PCA}}{Principal Component Analysis}
\newacronym{pcn}{{PCN}}{Parallel Coil Network}
\newacronym{primaldual}{{PD}}{Primal-Dual}
\newacronym{pd}{{PD}}{Proton Density}
\newacronym{pe}{{PE}}{Phase Encoding}
\newacronym{pet}{{PET}}{Positron Emission Tomography}
\newacronym{pg}{{PG}}{Proximal Gradient}
\newacronym{pi}{{PI}}{Parallel Imaging}
\newacronym{psf}{{PSF}}{Point Spread Function}
\newacronym{psnr}{{PSNR}}{Peak Signal-To-Noise Ratio}
\newacronym{rbf}{{RBF}}{Gaussian radial basis function}
\newacronym{relu}{{ReLU}}{Rectified Linear Unit}
\newacronym{rf}{{RF}}{Radio Frequency}
\newacronym{rmse}{{RMSE}}{Root-Mean-Squared-Error}
\newacronym{roi}{{ROI}}{Region Of Interest}
\newacronym{rss}{RSS}{Root-Sum-of-Squares}
\newacronym{sa}{{SA}}{Short Axis}
\newacronym{se}{{SE}}{Spin Echo}
\newacronym{sar}{{SAR}}{Specific Absorption Rate}
\newacronym{ssim}{{SSIM}}{Structural Similarity Index}
\newacronym{sense}{{SENSE}}{Sensitivity Encoding}
\newacronym{smash}{{SMASH}}{Simultaneous Acquisition of Spatial Harmonics}
\newacronym{sn}{{SN}}{Sensitivity Network}
\newacronym{snr}{{SNR}}{Signal-to-Noise Ratio}
\newacronym{spect}{{SPECT}}{Single Photon Emission Computed Tomography}
\newacronym{sqs}{{SQS}}{Separable Quadratic Surrogate}
\newacronym{stl}{{STL}}{Style Transfer Layer}
\newacronym{svd}{{SVD}}{Singular Value Decomposition}
\newacronym{tof}{{ToF}}{Time of Flight}
\newacronym{tgv}{{TGV}}{Total Generalized Variation}
\newacronym{te}{{TE}}{Echo Time}
\newacronym{tr}{{TR}}{Repetition Time}
\newacronym{tse}{{TSE}}{Turbo Spin Echo}
\newacronym{tv}{{TV}}{Total Variation}
\newacronym{us}{{US}}{Ultrasound}
\newacronym{vista}{{VISTA}}{Variable Density Incoherent Spatio-Temporal Acquisition}
\newacronym{vrs}{{VRS}}{Variable Density Random Sampling}
\newacronym{vn}{{VN}}{Variational Network}
\newacronym{vs}{{VS}}{Variable Splitting}
\newacronym{wgan}{{wGAN}}{Wasserstein Generative Adversarial Network}
\newacronym{zte}{{ZTE}}{Zero Echo Time}

\newcommand{\figI}{\mrmfigure{%
\includegraphics[width=\textwidth]{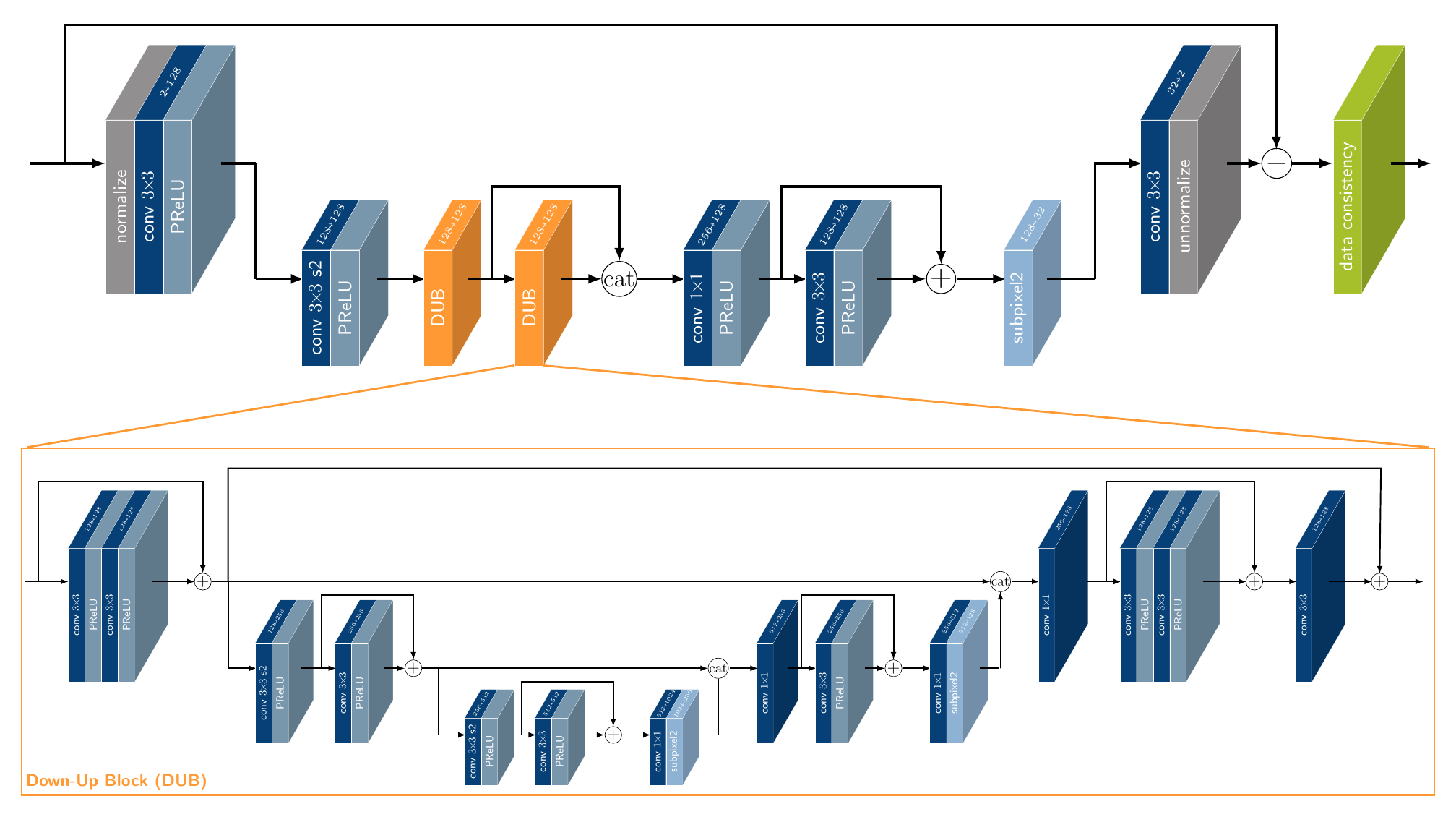}}%
{Structure of the Down-Up Network (DUN) block for MR image reconstruction with arbitrary \gls{dc} layers. This block is repeated in an iterative manner. Parameters over the single iterations can be shared or variable. Characteristic element is the concatenated Down-Up Block (DUB), consisting of numerous residual blocks at various scales.}%
{fig:figure1-dun}}

\newcommand{\figII}{\mrmfigure{%
\includegraphics[width=\textwidth]{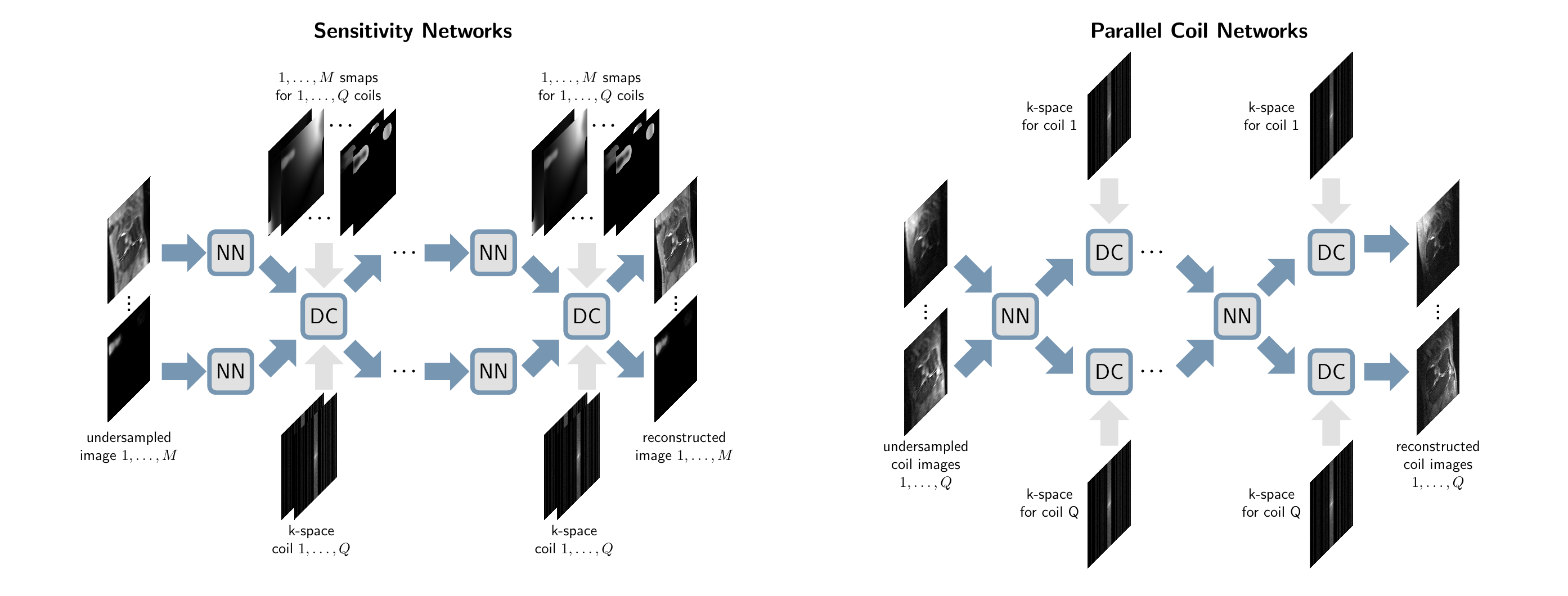}}%
{Structure of \acrfullpl{sn} and \acrfullpl{pcn}. While the combination of individual coil elements is done in the \gls{dc} layer in \glspl{sn} using explicit coil sensitivity maps, the \gls{nn} realizes the coil combination in \glspl{pcn}.}%
{fig:figure2-sn-pcn}}

\newcommand{\figIII}{\mrmfigure{%
\includegraphics[width=\textwidth]{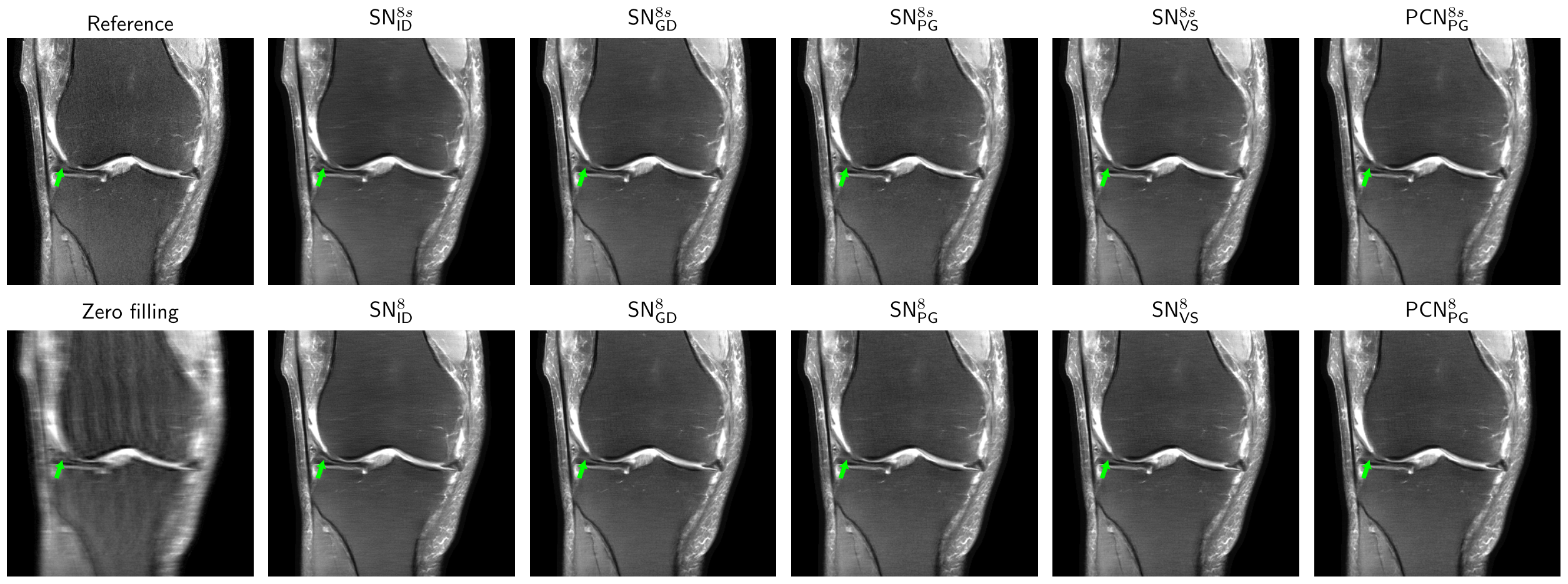}}%
{Comparison of different \gls{dc} terms for a coronal PDFS case and acceleration $R{=}4$ (\texttt{multicoil\_val/file1000990.h5}, slice 19). We observe a substantial difference between image enhancement and image reconstruction networks in terms of blurring and depicted anatomy. The lateral meniscus, indicated by the green arrow, shows a gap for image enhancement networks compared to reconstruction networks. We observe no substantial differences between different \gls{dc} layers. \sn{PG}{8s} contains more noise and appears slightly sharper than the other reconstructions.}
{fig:figure3-img_dc_PDFS_R4}}

\newcommand{\figIV}{\mrmfigure{%
\includegraphics[width=\textwidth]{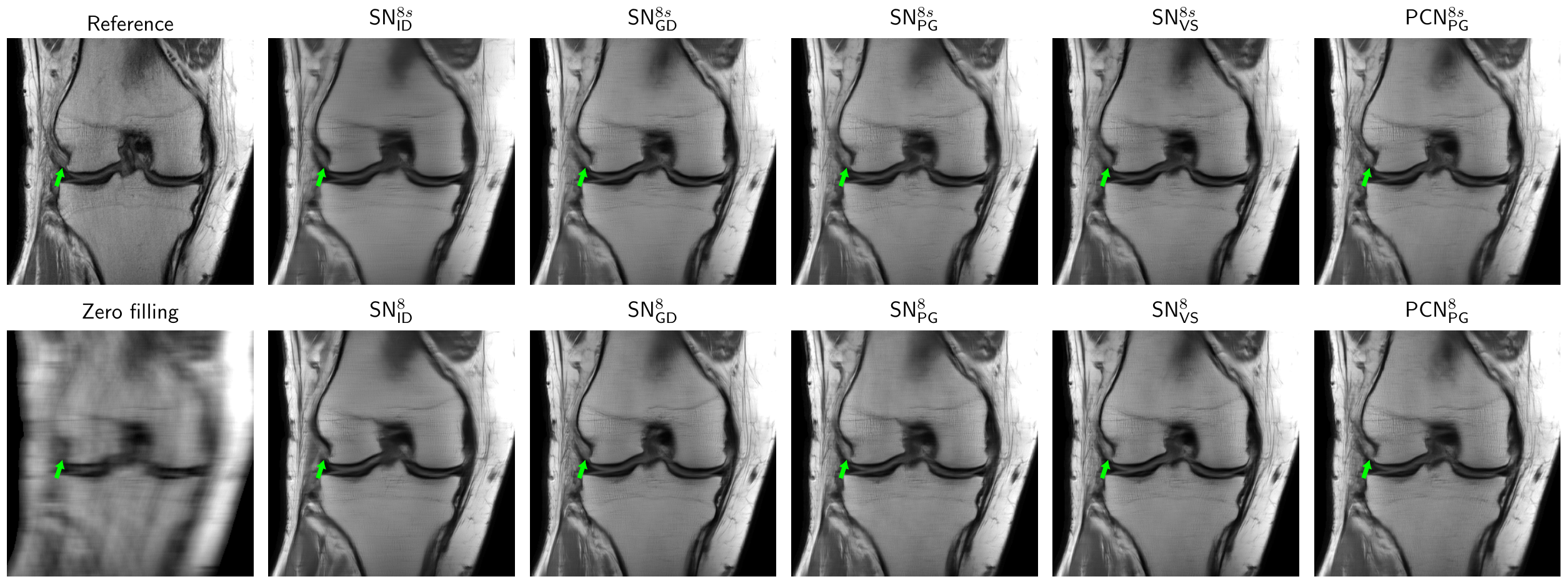}}%
{Comparison of different \gls{dc} terms for a coronal PD case and acceleration $R{=}8$ (\texttt{multicoil\_val/file1002351.h5}, slice 23). We observe a substantial difference between image enhancement and image reconstruction networks in terms of blurring and depicted anatomy. The area in the lateral aspect of the knee, indicated by the green arrows, shows a substantial deviation of anatomy for image enhancement networks, more prominent for \sn{ID}{8}. We observe no substantial differences between different \gls{dc} layers. \sn{PG}{8s} appears slightly sharper than the other reconstructions.}
{fig:figure4-img_dc_PD_R8}}

\newcommand{\figV}{\mrmfigure{%
\includegraphics[width=\textwidth]{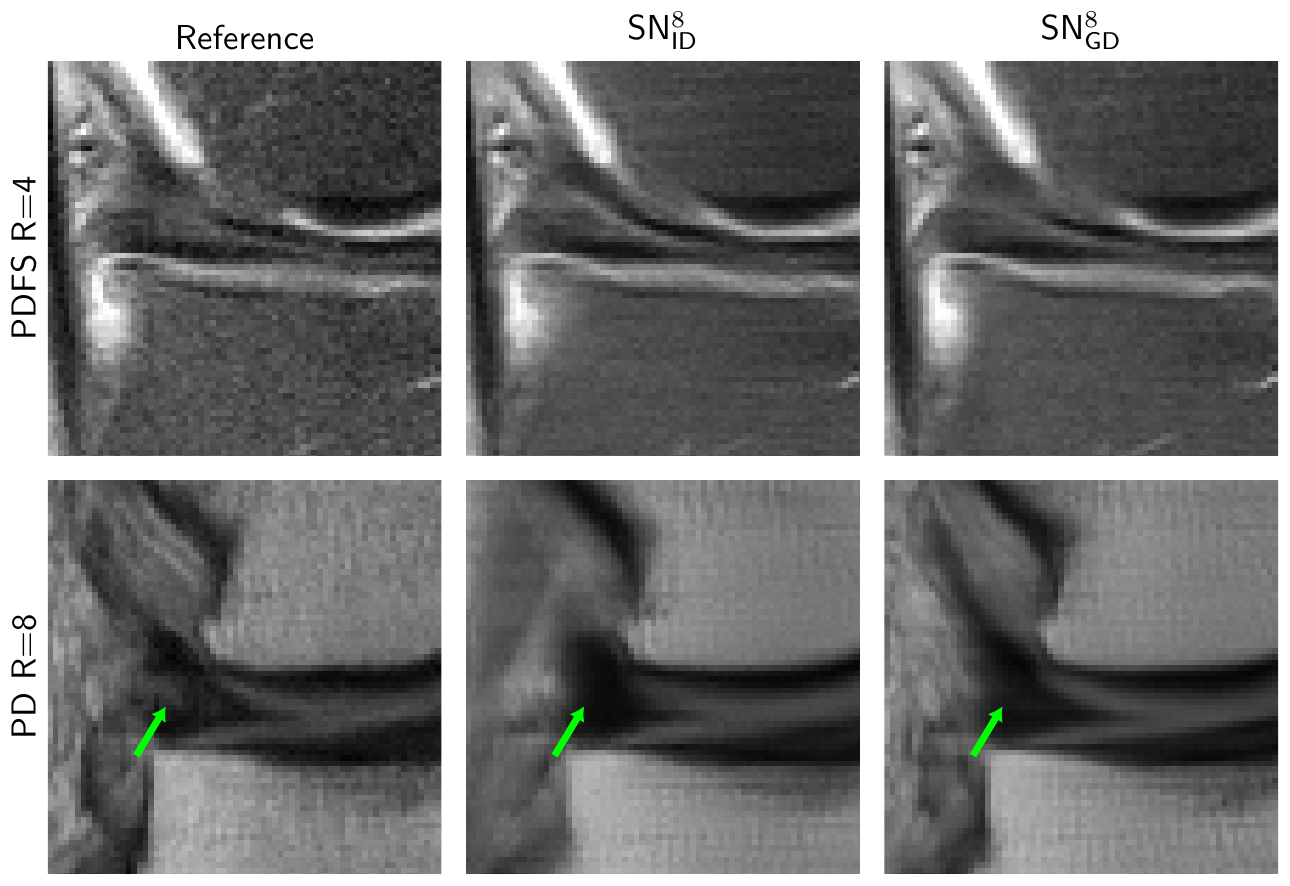}}%
{Detailed view of the depicted cases in \cref{fig:figure3-img_dc_PDFS_R4} and \cref{fig:figure4-img_dc_PD_R8} for image enhancement and image reconstruction. We clearly observe the deviations in anatomy for the image enhancement network in both contrasts and acceleration factors. The area in the lateral meniscus, indicated by the green arrow, is depicted neither in the image enhancement \sn{ID}{8} nor the reconstruction network \sn{GD}{8} for this high acceleration factor of $R{=}8$.}
{fig:figure5-img_dc_detail}}

\newcommand{\figVI}{\mrmfigure{%
\includegraphics[width=\textwidth]{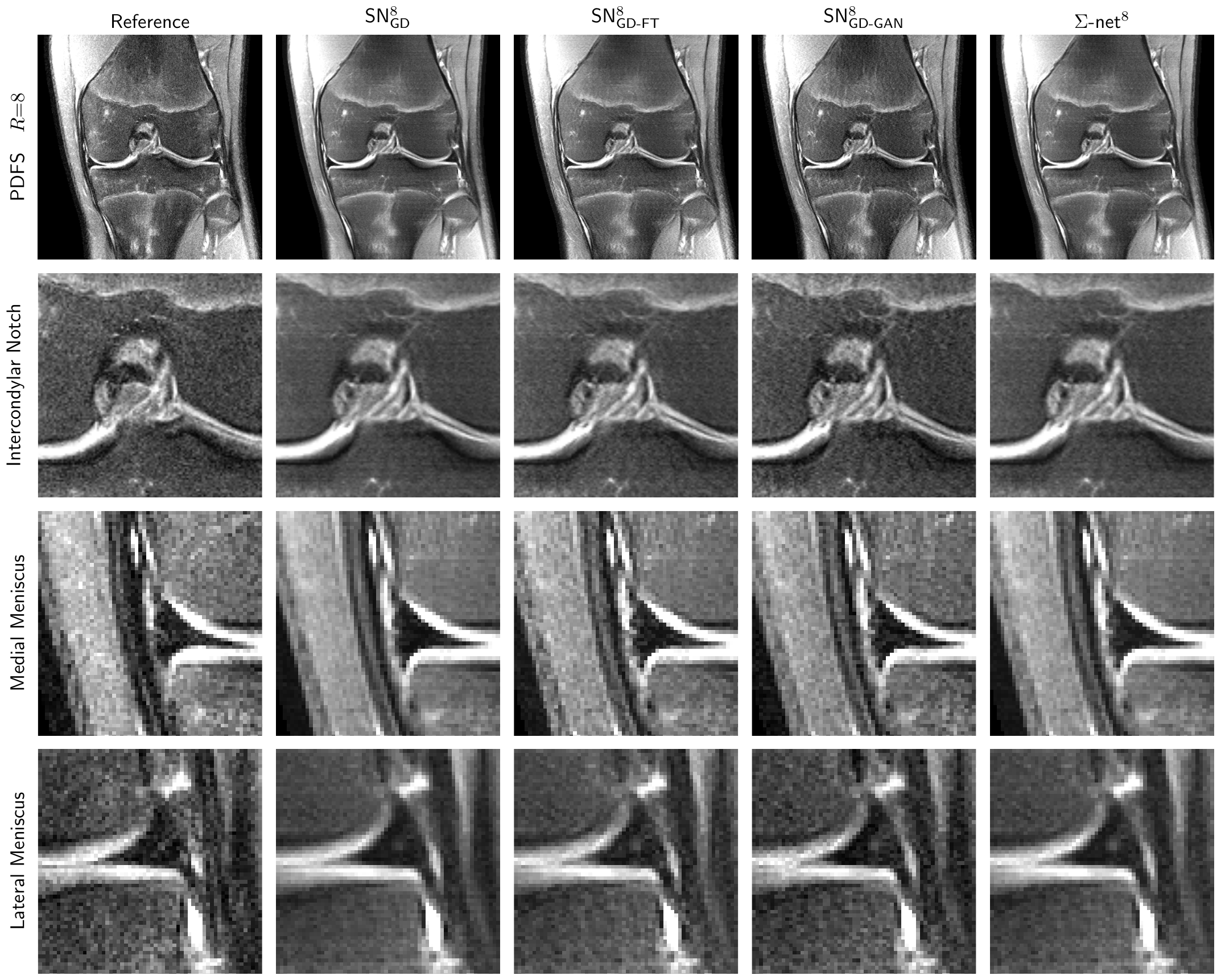}}%
{We show the influence of different learning schemes on a PDFS case (\texttt{multicoil\_val/file1000114.h5} for acceleration factor $R{=}8$ on the full knee, along with zoomed regions of the lateral and medial meniscus and the intercondylar notch. We observe that the image for \sn{GD}{8} is most blurred, and texture is increased for both \sn{GD-FT}{8} and \sn{GD-GAN}{8}. The \sn{GD-GAN}{8} result appears most textured. The $\Sigma$-net$^8$ results exploit the advantages of all included network architectures and show also improved texture compared to the \sn{GD}{8} network, while maintaining the quantitative scores as depicted in \cref{tab:table2-loss}.}
{fig:figure6-img_loss_PDFS_R8}}

\newcommand{\tabI}{%
		\mrmtable{\includegraphics[width=\textwidth]{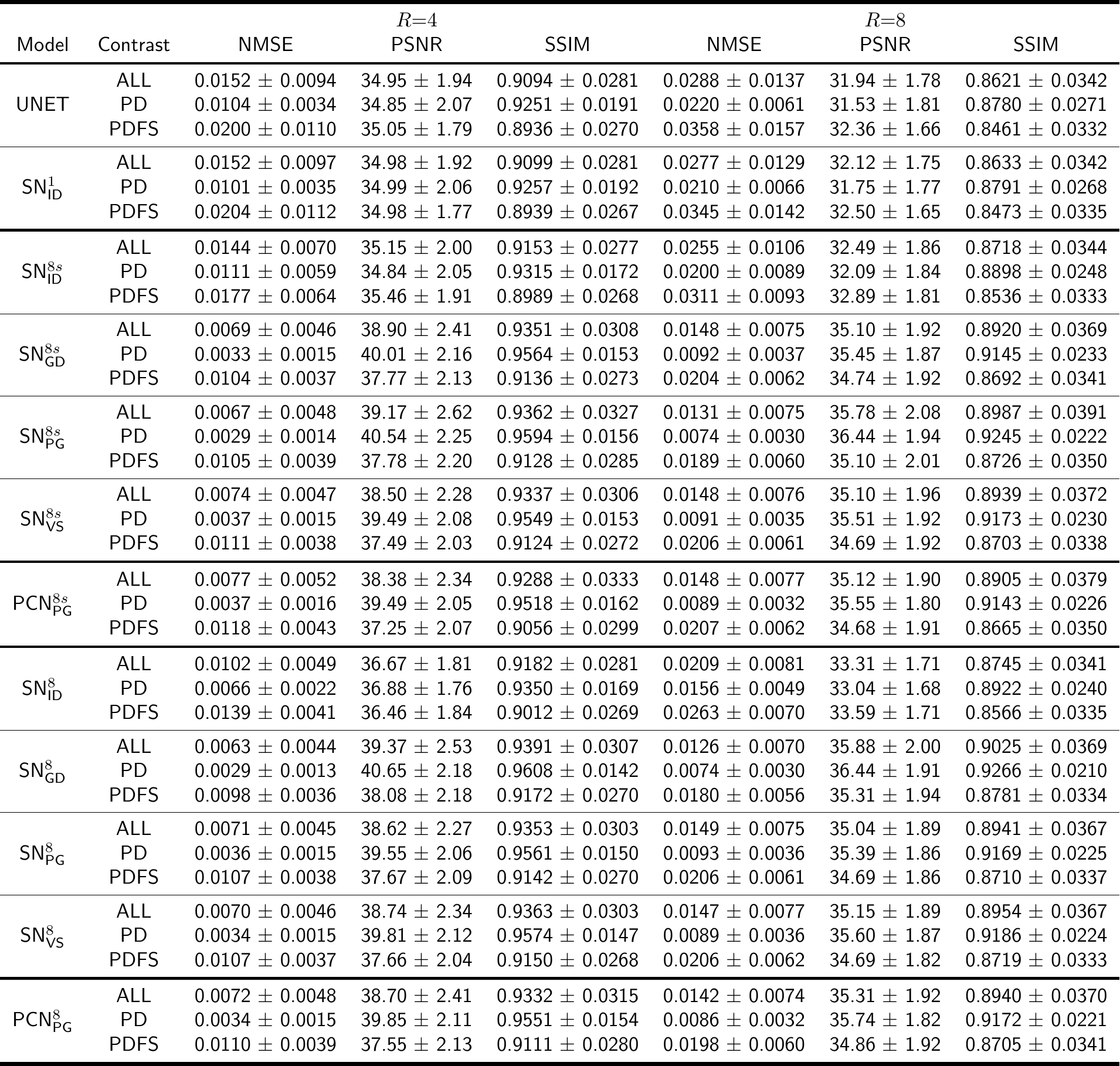}}
		{Comparison of \glspl{sn} and \glspl{pcn} for different \gls{dc} layers, number of cascades as well as shared and variable parameters. Reconstruction networks with any \gls{dc}, \ie GD, PG or VS, clearly outperform image enhancement networks.}{tab:table1-dc}}

\newcommand{\tabII}{%
		\mrmtable{\includegraphics[width=\textwidth]{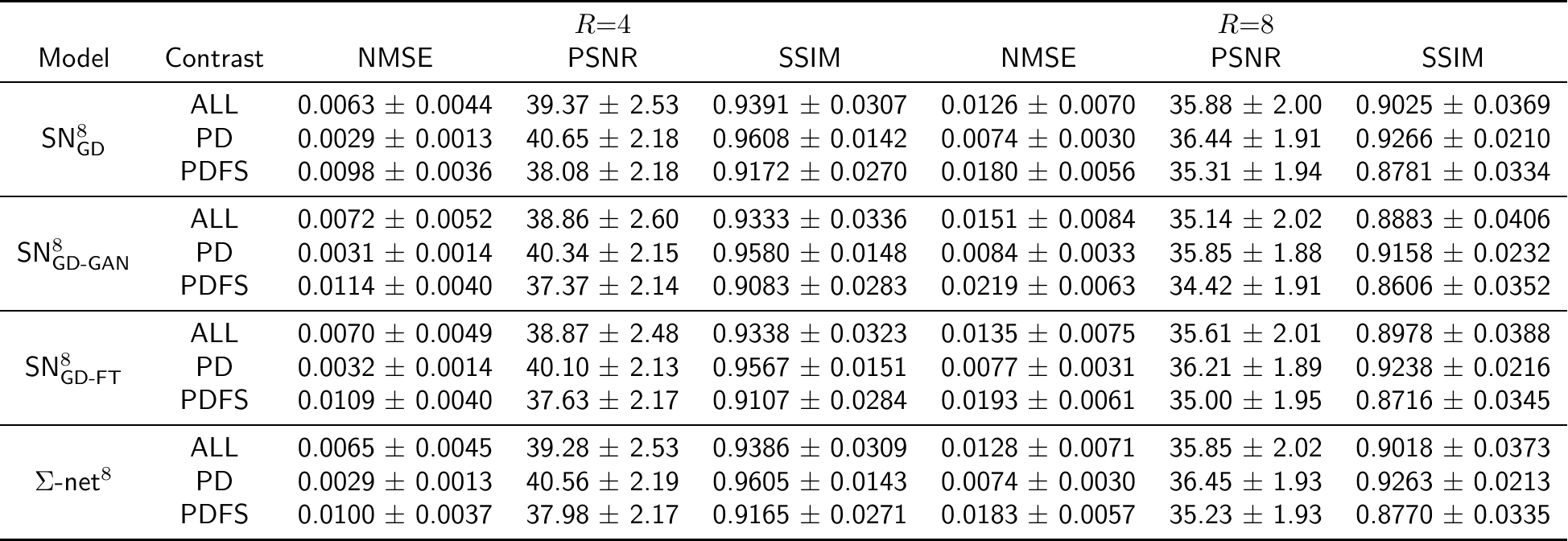}}
		{Influence of supervised learning with standard loss, adversarial loss and semi-supervised fine-tuning. By adding an adversarial loss, denoted by \gls{gan}, or using fine-tuning, the quantitative values drop for all cases.} {tab:table2-loss}}
		
\newcommand{\stabI}{%
	\mrmstable{\includegraphics[width=0.4\textwidth]{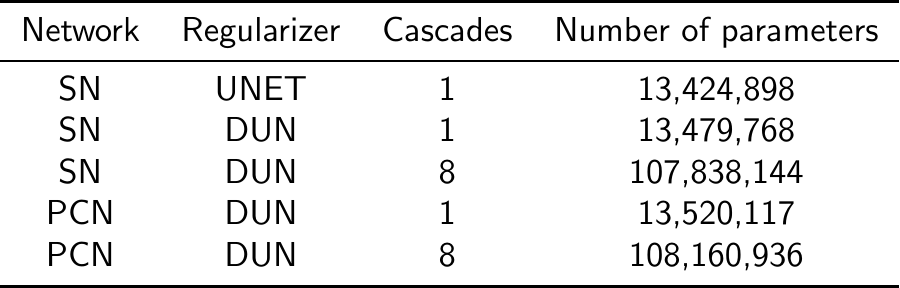}}
	 {Comparison of the number of parameters for \unet and \gls{dun}. The number of parameters for one cascades equals the number of parameters for arbitrary cascades and shared parameters. The \unet and \gls{dun} have approximately the same model capacity. The model capacity for \glspl{sn} and \glspl{pcn} are also comparable.}{stab:stable1-num-params}}
	
\newcommand{\sfigI}{\mrmsfigure{\includegraphics[width=\textwidth]{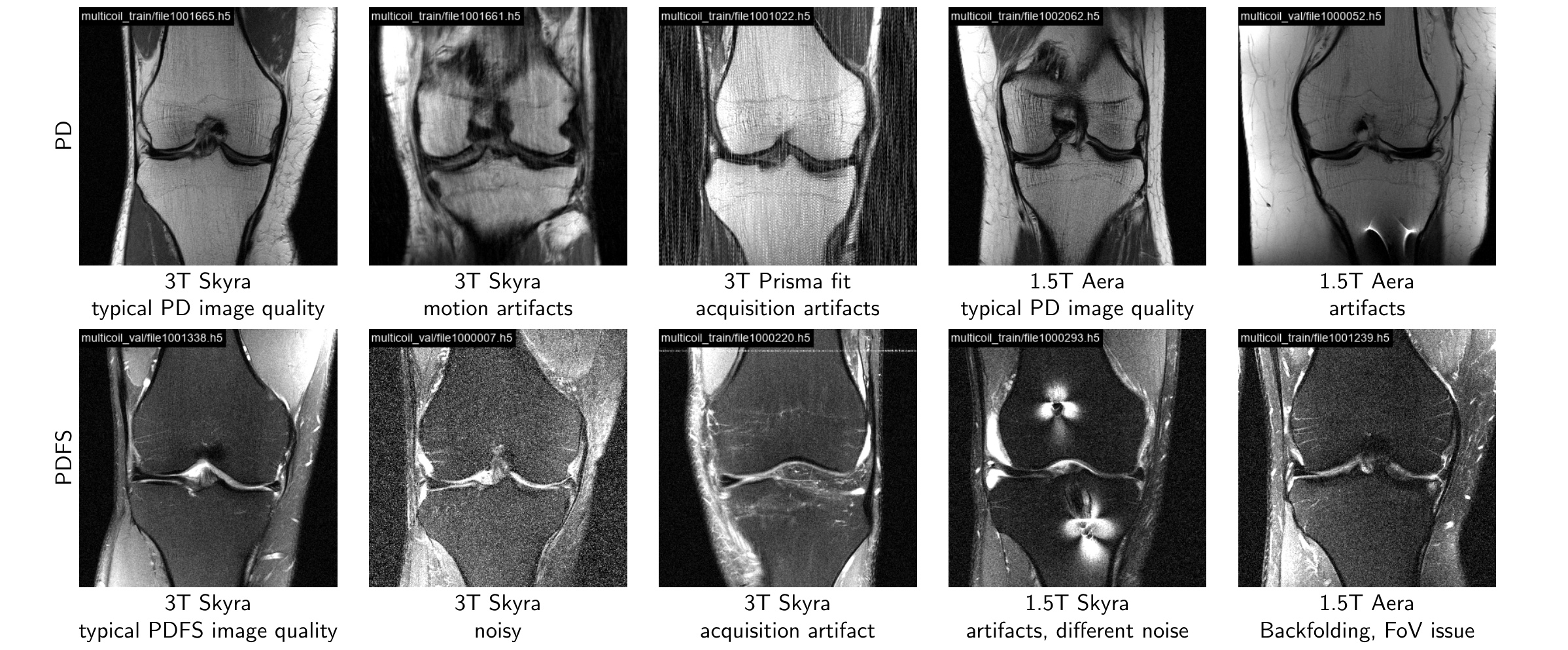}}
	{Examples of the heterogeneous \fastmri~\cite{Zbontar2018} training and validation set for coronal knee scans. The data have different number of \gls{pe} lines according to the acquisition.}{sfig:sfigure1-fastmri}}

\newcommand{\sfigII}{\mrmsfigure{\includegraphics[width=0.5\textwidth]{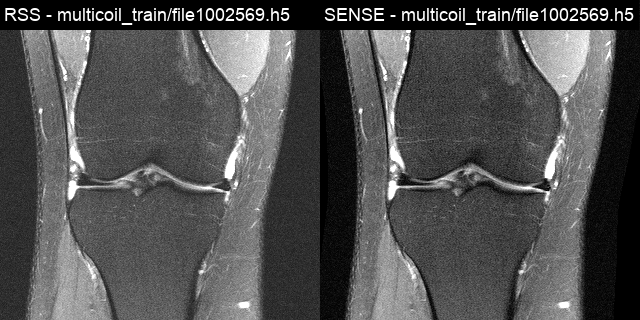}}
	{Comparison of fully-sampled \gls{rss} reconstruction (left), which is the reference of the \fastmri challenge, and the fully-sampled sensitivity-combined reconstruction (SENSE) (right). The \gls{ssim} between the two fully sampled scan is 0.6979, respectively. Although only foreground information is important, the background is also included in evaluation and has, however, a huge influence on the quantitative scores. Removing the background in both \gls{rss} and SENSE already yields an substantial increase of the \gls{ssim} to 0.9625.}{sfig:sfigure2-rss_vs_sense}}
	
\newcommand{\sfigIII}{\mrmsfigure{\includegraphics[width=\textwidth]{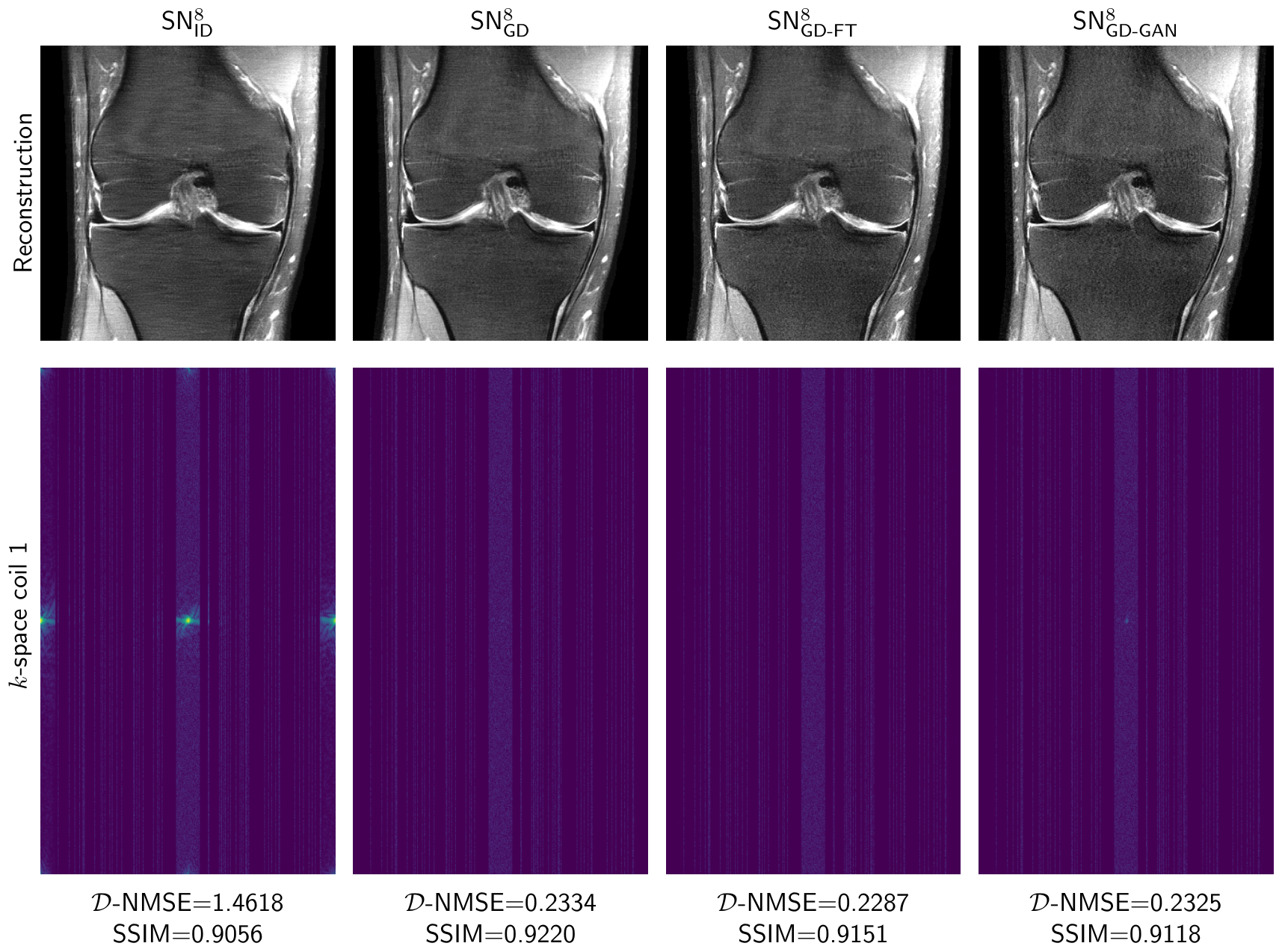}}
	{\texttt{multicoil\_val/file1000206.h5} slice 24 $R{=}4$: This figure depicts reconstruction results for different network architectures along with the dataterm argument $Ax-y$ of a selected coil. We observe huge deviations in \kspace for \sn{ID}{8}. The \kspace for \sn{GD}{8} and \sn{GD-FT}{8} appear similar. The \dnmse is lowest for \sn{GD-FT}{8}, indicating that the reconstruction adapted to the new \kspace data. Although the \dnmse is lower for  \sn{GD-GAN}{8} than for \sn{GD}{8} we observe a subtle, bright spot in the center of \kspace of \sn{GD-GAN}{8}, which might indicate a deviation from the ground truth for this particular slice. Note that despite better texture, the drop in \gls{ssim} of \sn{GD-FT}{8} from \sn{GD}{8} is expected, and is consistent with \cref{eq:semisupervised}.}{sfig:sfigure3-datanorm} }

\graphicspath{{./figures/}}

\title{\sigmanet: Systematic Evaluation of \\Iterative Deep Neural Networks for\\ Fast Parallel MR Image Reconstruction}

\runningtitle{\sigmanet: Systematic Evaluation of Iterative Deep Neural Networks for Fast Parallel MR Image Reconstruction}

\author{Kerstin~Hammernik$^{1*}$, Jo~Schlemper$^{1,2}$, Chen~Qin$^{1}$, Jinming~Duan$^{1,3}$, Ronald~M.~Summers$^{4}$ and Daniel~Rueckert$^{1}$}

\affiliation{$^1$ Department of Computing, Imperial College London, London, United Kingdom

\affilationspace
$^2$ Hyperfine Research Inc., Guilford, CT, USA

\affilationspace
$^3$ School of Computer Science, University of Birmingham, Birmingham, United Kingdom

\affilationspace
$^4$ NIH Clinical Center, MD, USA
}

\correspondence{
	$^*$Correspondence to: Kerstin Hammernik\\
	Department of Computing\\
	Imperial College London\\
	South Kensington Campus\\
	London SW7 2AZ, United Kingdom\\
	E-mail: k.hammernik@imperial.ac.uk\\
	
	\noindent Preliminary data for this manuscript were submitted to the \fastmri challenge (team \holykspace) and to ISMRM 2020.}

\grant{The work was funded in part by the EPSRC Programme Grant (EP/P001009/1) and by the Intramural Research Programs of the National Institutes of Health Clinical Center (1Z01 CL040004).}
\numWordsAbstract{206}
\numWordsBody{Approx. 5000}
\numFigures{8}
\numCitations{89}
\numSupporting{4}

\keywords{Parallel~Imaging, Iterative~Image~Reconstruction, Deep~Learning, Down-Up~Networks, Data~Consistency, Ensembling}

\begin{document}
	
\maketitle
	
\begin{abstract}
	\paragraph{Purpose}
	To systematically investigate the influence of various data consistency layers, (semi-)supervised learning and ensembling strategies, defined in a \sigmanet, for accelerated parallel MR image reconstruction using deep learning.
	
	\paragraph{Theory and Methods}
	MR image reconstruction is formulated as learned unrolled optimization scheme with a Down-Up network as regularization and varying data consistency layers. The different architectures are split into sensitivity networks, which rely on explicit coil sensitivity maps, and parallel coil networks, which learn the combination of coils implicitly.
	Different content and adversarial losses, a semi-supervised fine-tuning scheme and model ensembling are investigated.

	\paragraph{Results}
	Evaluated on the \fastmri multicoil validation set, architectures involving raw \kspace data outperform image enhancement methods significantly.
	Semi-supervised fine-tuning adapts to new \kspace data and provides, together with reconstructions based on adversarial training, the visually most appealing results although quantitative quality metrics are reduced.
	The \sigmanet ensembles the benefits from different models and achieves similar scores compared to the single \sota approaches.
	
	\paragraph{Conclusion}
	This work provides an open-source framework to perform a systematic wide-range comparison of \sota reconstruction approaches for parallel MR image reconstruction on the \fastmri knee dataset and explores the importance of data consistency.
	A suitable trade-off between perceptual image quality and quantitative scores are achieved with the ensembled \sigmanet.
	
\end{abstract}

\clearpage

\section{Introduction}

\gls{pi}~\cite{Sodickson1997, Pruessmann2001, Griswold2002, Uecker2014} forms the foundation of accelerating data acquisition in \gls{mri}, which is tremendously time-consuming.
In the last decade, \gls{pi} combined with \gls{cs} techniques have resulted in substantial improvements in acquisition speed and image quality~\cite{Lustig2006,Liang2009,Murphy2012,Shin2014,Haldar2013,Haldar2016,Jin2016,Block2007} and many advanced regularization techniques have been proposed~\cite{Knoll2012,Ravishankar2012,Caballero2014}.
Although \gls{pi}-\gls{cs} can achieve \sota performance, designing effective regularization schemes and tuning of hyper-parameters are non-trivial.
Starting in 2016, deep learning algorithms have become extremely popular and effective tools in data-driven learning of inverse problems and have enabled progress beyond the limitations of \gls{cs}~\cite{Hammernik2018, Wang2016ab, Han2018, Zhu2018, Schlemper2018a, Yang2016}.

Current research of deep learning for (parallel) \gls{mri} reconstruction focuses on two major branches.
The first branch deals with the design of effective network architectures and loss functions.
To this end, many different architectures were proposed, exploiting different content losses such as the $\ell_1$ norm and \gls{ssim}~\cite{Zhao2015,Hammernik2017f} and adversarial losses based on \glspl{gan}~\cite{Yang2017,Liu2017,Quan2017,Seitzer2018,Mardani2018}.
The second branch deals with the question of how we can incorporate the rich domain knowledge of \gls{mri} in the reconstruction models.
Various types of domain knowledge have been explored and many authors have exploited \gls{dc}, unique to singlecoil and multicoil imaging, iterative reconstruction approaches~\cite{Mardani2018, Aggarwal2018,Cheng2018,Kwon2017a,Hammernik2018}, $k$-space learning~\cite{Han2018a,Eo2018,Souza2018a,Akcakaya2019}, complex geometry~\cite{Virtue2017, Wang2019a} and dedicated applications such as multi-contrast reconstruction~\cite{Sun2019,Polak2019}, dynamic \gls{mri}~\cite{Qin2017,Schlemper2018b,Qin2019,Hauptmann2018a,Jin2019,Kofler2019} and \gls{mr} fingerprinting~\cite{Cohen2018,Virtue2017}.
For further references in this field, we refer the interested reader to survey papers~\cite{Knoll2019, Hammernik2018b, Liang2019, Lundervold2019}.

Despite the tremendous volume of research, there remain several challenges in the field of learning \gls{mr} image reconstruction.
Learning-based approaches are often evaluated in dedicated research settings with small, homogeneous datasets, which are often not available to the public.
Hence, it is often non-trivial to draw reliable conclusions from the reported findings because the results highly depend on the dataset and hyper-parameter tuning, which make many results non-reproducible.
As such, it is still largely unclear which types of approaches are more effective and if they are effective in general or only effective for the application studied for a specific experimental environment.

The \fastmri knee dataset~\cite{Zbontar2018} is the first publicly available dataset that mitigates the issue of small, homogeneous datasets for \gls{mr} image reconstruction, and was already used for evaluation in~\cite{Putzky2019,Souza2018a,Bahadir2019,Sriram2019,Wang2019b}.
The \fastmri knee dataset consists of almost 1,500 clinical knee datasets from various Siemens \gls{mr} scanners at different field strengths and contains fully-sampled coronal \gls{pd} scans with and without \gls{fs}, which can then be retrospectively undersampled.
Examples of challenging and heterogeneous \fastmri datasets are illustrated in \cref{sfig:sfigure1-fastmri}.
The \fastmri dataset is accompanied by the \fastmri challenge, which provides a great opportunity to push the limits of acquisition speed and compare a variety of approaches developed by the community on a common basis.

The evaluation of the \fastmri challenge relies on a \gls{rss} reconstruction of the fully-sampled \kspace data.
However, \gls{rss} reconstruction provides a bias towards approaches which do not depend on explicit coil sensitivity maps, which is evident when we compare the \gls{rss} and a sensitivity-combined reference image qualitatively and quantitatively, as shown in \cref{sfig:sfigure2-rss_vs_sense}.
The bias is due to the fact that the \gls{rss} combination of the individual coil channels accumulates the magnitude components of the noise, which causes a bias in low signal intensity regions and in the background.
Additionally, the noise does not remain Gaussian any more as the magnitude of the noise adds up in the final image, which is not the case if the images are combined with explicit coil sensitivity maps.
This forces the network to learn the scan-specific noise bias, which is undesirable.
Therefore, in this work, we consider sensitivity-combined reconstruction as a ground truth to appropriately evaluate the efficacy of multicoil approaches that might involve explicit coil sensitivity maps.
Note that pre-whitening~\cite{Roemer1990,Robson2008} of the data might reduce this bias effect in the \gls{rss} reconstruction and also in the sensitivity-combined reconstruction, as both the noise influence of possibly broken coil elements as well as correlations between coil elements might be compensated.
%

%

The aim of this work is to bridge the gap of aforementioned challenges that we have observed in deep learning for parallel \gls{mr} image reconstruction.
First, we introduce the concept of \glspl{sn}, relying on explicit coil sensitivity maps, and \glspl{pcn}, that learn sensitivity maps implicitly.
Second, we perform a large-scale evaluation of different networks with varying \gls{dc} layers and a common \gls{dun} as regularization, which allows us to directly compare image enhancement and image reconstruction approaches with the same model capacity.
Third, we study the influence of supervised learning, including both content and adversarial losses, semi-supervised fine-tuning and model ensembling, defined as \sigmanet.
These three concepts represent amongst others many \sota approaches including~\cite{Jin2017,Lee2017a,Zbontar2018,Hammernik2018,Schlemper2018a,Aggarwal2018,Mardani2018,Schlemper2019,Duan2019}, enabling a systematic comparisons.
All experiments are performed on the \fastmri multicoil knee dataset~\cite{Zbontar2018}, where a fully-sampled sensitivity-combined reconstruction is used as a re-defined reference for the dataset, which we denote as SENSE target throughout our manuscript.
%
A first evaluation of \sigmanet on the \gls{rss} references was targeted in our \fastmri challenge submission~\cite{Schlemper2019a} with our team \holykspace, where \sigmanet is among the top three performing methods on the \fastmri multicoil test and challenge dataset.

\section{Theory}
\subsection{Learning Unrolled Optimization}
Accelerated \gls{mr} image reconstruction aims at recovering a reconstruction $x\in\mathbb{C}^{N_x}$ from a set of undersampled \kspace measurements $y \in \mathbb{C}^{N_y}$ which are corrupted by additive Gaussian noise $n\in\mathbb{C}^{N_y}$ following
\begin{align}
    y = Ax + n.\label{eq:ip}
\end{align}
This inverse problem involves a linear forward operator $A:\mathbb{C}^{N_x}\to\mathbb{C}^{N_y}$ modelling the \gls{mr} physics.
Here, $N_x$ and $N_y$ define the dimensions of the reconstruction $\x$ and the \kspace data $y$ according to the underlying multicoil or singlecoil problem. These dimensions and the linear forward operators will be specified separately in the later sections for \glspl{sn} and \glspl{pcn}.

An approximate solution to the inverse problem in \cref{eq:ip} can be obtained by learning a fixed iterative reconstruction scheme~\cite{Hammernik2018,Aggarwal2018,Schlemper2018b,Qin2017,Duan2019}.
We define the fixed unrolled algorithm for \gls{mr} image reconstruction as
\begin{align}
    x^\tphalf &= x^\t - f_{\theta^\t}(x^\t),\\
    x^\tp &= g(x^\tphalf, y, A)\label{eq:dc}
\end{align}
for $0 \leq \t < \T$.
Here, $f_\theta$ represents a \gls{nn}-based reconstruction block with trainable parameters $\theta$, $g$ denotes a \gls{dc} layer and $T$ is the number of fixed iterations, also termed cascades~\cite{Schlemper2018b} or stages~\cite{Hammernik2018} in literature.
The reconstruction block in $f_\theta$ has the form of an encoding-decoding structure, which can be realized by any type of \glspl{cnn}, \unet architectures~\cite{Ronneberger2015} or is motivated by variational methods~\cite{Hammernik2018}.
In this work, we show the efficiency of a \gls{dun}~\cite{Yu2019} over \unet for \gls{mr} image reconstruction.
The \gls{dc} can be realized in various ways, similar to choosing an optimizer for solving inverse problems.
In this work, we compare \gls{gd}~\cite{Hammernik2018}, \gls{pg}~\cite{Schlemper2018b,Aggarwal2018} and \gls{vs}~\cite{Duan2019} schemes.
Both \glspl{dun} and \gls{dc} will be presented in the following sections.

\subsubsection{\acrfullpl{dun}}
\acrfullpl{dun}~\cite{Yu2019} have shown many advantages compared to \unet, which will be shortly outlined, and they have also achieved top performances in the NTIRE 2019 challenge on real image denoising~\cite{Abdelhamed2019}.
%
%
The \gls{dun} as shown in \cref{fig:figure1-dun} first downsamples the image by convolutions with stride 2 and then performs analysis on a coarser scale.
The method then applies multiple \glspl{dub}, similar to a \unet but with improved residual connections. 
The outputs of the \glspl{dub} are concatenated and further analyzed by a residual convolution/activation block, followed by sub-pixel convolutions \cite{Shi2016a}.
Several things can be noted:
First, \glspl{dun} are more memory efficient than conventional networks because the \gls{dun} performs the bulk of the computations on a coarser scale~\cite{Yu2019}.
This finding was supported by comparing the memory footprint on the \gls{gpu} of \unet and \gls{dun} with the same model capacity and number of base features $n_f{=}64$.
We observe that the \gls{dun} only requires 72\% of \gls{gpu} memory compared to \unet.
Note that shifting the computation to a coarser scale has not shown any limitation on analysis at the finest scale \cite{Yu2019, Zhang2018d}.
In fact, the \gls{dun} allows one to increase the number of features at coarse levels, which improves the expressiveness of the model.
Second, the iterative down-up structures have shown to be efficient in propagating information at different scales, resulting in improved performance for super-resolution tasks \cite{Haris2018}.
Third, the \gls{dun} relies on sub-pixel convolution, which performs superior in terms of expressiveness and computational efficiency over upsampling convolution \cite{Shi2016}. 

In our \gls{dun}, presented in \cref{fig:figure1-dun}, we first perform complex-valued whitening~\cite{Trabelsi2017} to normalize the image and perform the learning in normalized space.
The complex-valued image is split into two channels, containing the real and imaginary parts.
We start with an initial number of filters $n_f{=}64$, which is doubled after each downsampling and used 2 \glspl{dub}.
In the upsampling path, the number of features is doubled prior to sub-pixel shuffling to preserve the number of features.
Each downsampling and upsampling is performed with stride 2.
Finally, the reconstruction is unnormalized using the initial normalization parameters.
Note that we did not use batch-normalization as this has been shown to have unexpected aliasing for image restoration tasks \cite{Wang2018b}.
While in other works it is common to use dense connections~\cite{Huang2017,Zhang2018c}, we did not utilise these in our network either due to their high memory requirement.

\subsubsection{\acrfull{dc}}
Using \gls{dc}, modelled by $g$ in \cref{eq:dc}, we have the possibility to impose prior knowledge about the \gls{mr} acquisition process to the learning-based reconstruction scheme.
Assuming Gaussian noise $n$ in the measurement data $y$, the $\ell_2$ norm is a suitable similarity measure
\begin{align*}
    \data{Ax,y} = \frac{\lambda}{2}\norm[2]{Ax - y}^2.
\end{align*}
As \gls{dc} follows a regularization by a \gls{nn} in image domain, the parameter $\lambda$ balances the impact of the \gls{dc} term $\data{Ax,y}$.
When including prior knowledge about the acquisition process, we have various possibilities.
The simplest case is to perform a gradient step~\cite{Hammernik2018} related to the \gls{dc} term $\data{Ax,y}$
\begin{align}
    g_{\textsc{gd}}(x^\tphalf) = x^\tphalf - \lambda^t \AT (Ax^\tphalf - y),\label{eq:gd}
\end{align}
where $\AT$ denotes the adjoint operator of $A$.
Instead of alternating \gls{gd}, we can use a \gls{pg} scheme, where the \gls{dc} is modelled by the proximal mapping
\begin{align}
g_{\textsc{pg}}(x^\tphalf) = \argmin_x \frac{1}{2} \norm[2]{x - x^\tphalf}^2 + \frac{\lambda}{2} \norm[2]{Ax - y}^2.\label{eq:prox}
\end{align}
This is especially feasible if the proximal mapping is easy to compute and a closed-form solution exists.
If no closed-form solution exists, as this is typically the case for parallel MRI involving coil sensitivity maps, the proximal mapping can be solved numerically using a conjugate gradient optimizer~\cite{Aggarwal2018}.

For inverse problems where the inverse of $A$, hence, the proximal mapping in \cref{eq:prox} is intractable to compute, the operations involved in $A$ can be split using an additional variable as proposed for sensitivity-weighted parallel MRI in~\cite{Duan2019}.
To review \gls{vs}, we first introduce the sensitivity-weighted multi-coil operator for the $q^{\text{th}}$ coil as $A_q=M\FT C_q$.
The operator $C_q:\mathbb{C}^{N_x}\to\mathbb{C}^{N_x}$ applies the $q^{\text{th}}$ pre-computed coil sensitivity map to $x$, for $q=1,\ldots,Q$. This is followed by a \gls{ft} $\FT: \mathbb{C}^{N_x}\to\mathbb{C}^{N_x}$.
The operator $M: \mathbb{C}^{N_x}\to\mathbb{C}^{N_y}$ realizes the Cartesian sampling pattern and masks out \kspace lines that where not acquired.
\gls{vs} divides the problem defined in \cref{eq:prox} in two sub-problems by using a coil-wise splitting variable $z_q\in\mathbb{C}^{N_x}$
\begin{align*}
    z_q^\tp &= \argmin_{z_q} \frac{\lambda}{2} \sum\limits_{q=1}^{Q}\norm[2]{M\FT z_q - \y_q}^2 + \frac{\alpha}{2}\sum\limits_{q=1}^{Q}\norm[2]{z_q - C_qx^\tphalf}^2\\
    g_{\textsc{vs}}(x^\tphalf) &= \argmin_x \frac{\alpha}{2}\sum\limits_{q=1}^{Q}\norm[2]{z_q^\tp - C_qx}^2 + \frac{\beta}{2} \norm[2]{x - x^\tphalf}^2,
\end{align*}
where $\alpha > 0$ and $\beta > 0$ balance the influence of the soft constraints.
Solving these sub-problems yields the following closed-form solution
\begin{align*}
    z_q^\tp &= \IFT \left(\left(\lambda M^* M + \alpha I\right)^{-1}\left(\alpha \FT C_q x^\tphalf + \lambda M^* y_q\right)\right)\\
    g_{\textsc{vs}}(x^\tphalf) &= \left(\beta I + \alpha \sum\limits_{q=1}^{Q} C_q^*C_q \right)^{-1} \left( \beta x^\tphalf + \alpha \sum\limits_{q=1}^{Q}C_q^* z_q^\tp \right).
\end{align*}
Here, $I$ denotes the identity matrix and $^*$ the adjoint operation.

All presented \gls{dc} layers, \ie \gls{gd}, \gls{pg} and \gls{vs} ensure soft \gls{dc} to the measurement data $y$, representing image reconstruction networks.
By setting $\lambda{=}0$ in \cref{eq:gd} we obtain
%
\begin{align*}
    g_{\textsc{id}}(x^\tphalf) = x^\tphalf.
\end{align*}
This represents an image enhancement network, where the influence of \gls{dc} is omitted.

\subsection{\acrlongpl{sn} and \acrlongpl{pcn}}
We investigate two types of architectures for parallel \gls{mr} image reconstruction: \acrfullpl{sn} and \acrfullpl{pcn}, visualized in \cref{fig:figure2-sn-pcn}.
The major difference between these network architectures is their way of (weighted) coil combination.

For \glspl{sn}, the coil combination is defined with the operator $A:\mathbb{C}^{N_x}\to\mathbb{C}^{N_y}$ using explicit coil sensitivity maps as in~\cite{Hammernik2018}.
To overcome \gls{fov} issues in the SNs, we use an extended set of $M{=}2$ coil sensitivity maps according to~\cite{Uecker2014}, hence, reconstructing $x=[x_1,x_2]$.
The dimensions are given as $N_x=N_{\textsc{fe}}\cdot N_{\textsc{pe}}\cdot M$ and $N_y=N_{\textsc{fe}}\cdot N_{\textsc{pe}}\cdot Q$, where $N_{\textsc{fe}}$ and $N_{\textsc{pe}}$ denote the number of \gls{fe} and \gls{pe} lines, respectively.
In the case of \glspl{sn}, the \gls{nn} $f_\theta:\mathbb{C}^{N_x}\to\mathbb{C}^{N_x}$ has two complex-valued input and output channels.

In contrast, \glspl{pcn} reconstruct individual coil images $x=[x_1,\ldots,x_Q]$ for $Q$ coils, hence, $N_x=N_y=N_{\textsc{fe}}\cdot N_{\textsc{pe}}\cdot Q$.
The operator $A:\mathbb{C}^{N_x}\to\mathbb{C}^{N_y}$ simply performs coil-wise \glspl{ft} and \kspace masking and results in the same \gls{dc} term as in~\cite{Schlemper2018b} for single-coil \gls{mr} imaging.
The \gls{nn} $f_\theta:\mathbb{C}^{N_x}\to\mathbb{C}^{N_x}$ of \glspl{pcn} has $Q$ complex-valued input and output channels and learns weighted coil combination implicitly.

For both \glspl{sn} and \glspl{pcn}, the final reconstruction $x_{\text{rec}}$ is obtained by a \gls{rss} combination of the individual network output channels of $x^T$.
The \gls{nn} is realized using a \gls{dun}~\cite{Yu2019} as described in the previous section.
For the \gls{pcn}, only \gls{pg} networks will be investigated as no coil sensitivity maps are involved and a closed-form solution for the proximal mapping \cref{eq:prox} exists.

\subsection{Supervised Learning}
The networks were trained using a combined $\ell_1$ and \gls{ssim}~\cite{Zhao2015,Hammernik2017f} content loss $\ell_\text{base}$ between the reference $x_{\text{ref}}$ and the reconstruction $x_\text{rec}{=}x^T$
\begin{align*}
    \ell_\text{base}(x_\text{rec}, x_\text{ref}) = 100 - \text{SSIM}_\%(m \odot |x_\text{rec}|, m \odot |x_\text{ref}|) + \gamma_{\ell_1} \ell_1(m \odot |x_\text{rec}|, m \odot |x_\text{ref}|).
\end{align*}
where $\odot$ is the pixel wise product and $|\cdot|$ denotes the \gls{rss} reconstruction to combine the individual output channels.
The $\text{SSIM}_\%$ is given in percent only during training, to increase numerical stability.
This loss formulation also involves a binary foreground mask $m$ to focus the network training on the image content and not on the background.
The parameter $\gamma_{\ell_1}{=}10^{-3}$ is chosen empirically to match the scale of the two losses.

In addition to the content loss $\ell_\text{base}$, we investigate the impact of an adversarial loss based on  \gls{lsgan}.
%
%
%
Specifically, we adopt the \gls{lsgan} formulation as proposed in \cite{Mao2017}, and combine it with $\ell_{\text{base}}$, yielding the following two objective functions, which are optimized in an alternating manner,
\begin{align*}
\min_\eta \hspace{2pt} &\frac{1}{2} {\mathbb{E}}_{x_\text{ref} \sim p_{x_\text{ref}}}\big[\left(h_\eta(m \odot |x_\text{ref}|)-1\right)^2]+\frac{1}{2} {\mathbb{E}}_{x^0 \sim p_{x^0}}\big[\left(h_\eta(m \odot |x_{\text{rec},\theta}|\right)^2]
\\ 
\min_\theta \hspace{2pt} & \frac{1}{2} {\mathbb{E}}_{x^0 \sim p_{x^0}}\big[\left(h_\eta(m \odot |x_{\text{rec},\theta}|)-1\right)^2]+\gamma_{\text{base}}\ell_\text{base}(m \odot |x_{\text{rec},\theta}|, m \odot |x_\text{ref}|).
\end{align*}
Here, $h_\eta$ defines a discriminator network with trainable parameters $\eta$  which has the task to identify if the input is a fully-sampled image $x_\text{ref}$ or the generated reconstruction image $x_{\text{rec},\theta}$.
Note that we add $\theta$ in the subscript here to clarify the dependency on trainable parameters $\theta$ of the reconstruction network.
Furthermore, $p_{x_\text{ref}}$ and $p_{x^0}$ are distributions over reference data $x_\text{ref}$ and undersampled zero-filling reconstructions $x^0$, respectively.
Here, $x^0$ serves as initialization of the reconstruction network to generate $x_{\text{rec},\theta}=x^T$.
For the discriminator, we used the same architecture as for the super-resolution \gls{gan} proposed in~\cite{Ledig2017}.
The hyper-parameter $\gamma_{\text{base}}$ defines the trade-off between the effects of the content and adversarial losses, and is chosen as 0.1 empirically.

\subsection{Semi-Supervised Fine-Tuning}
A major drawback of supervised learning approaches is their tendency to produce overly smooth reconstructions.
Pixel-based losses such as the $\ell_1$ or $\ell_2$ norm do not have the capability to account for fine details and texture in the images and are known for their averaging behaviour~\cite{Ledig2017}.
Even patch-based losses such as \gls{ssim} do not have the capability to adapt to the characteristics of \gls{mr} images.
Hence, supervised learning produces the best reconstruction to account for a wide range of given training data from various scanners, contrasts and noise levels, according to the underlying loss function.
However, if new data are presented which have different characteristics, the trained model might not produce appealing results.
To overcome these issues, we propose a semi-supervised fine-tuning approach, motivated by~\cite{Ulyanov2018,Mataev2019}.
We consider the problem
\begin{align}
    \min_\theta \frac{1}{2}\norm[2]{Ax_{\text{rec},\theta} - y}^2 + \gamma_{\text{prior}} \max\left(1 - \text{SSIM}_\%(m \odot |x_{\text{rec},\theta}|, m \odot|x_{\text{prior}}|) - \gamma_{\text{th}}), 0\right)^2. \label{eq:semisupervised}
\end{align}
The aim is to adapt the network parameters $\theta$ to new \kspace data, ensured by the first \gls{dc} term.
Instead of using a latent noise vector as in~\cite{Ulyanov2018}, we already have a pre-trained reconstruction network involving available prior knowledge, which slightly needs to be adapted to the new data.
Hence, $\theta$ is initialized by the learned parameters of a supervised learning approach.
To avoid overfitting to the \kspace data, we add a regularization term based on the supervised model reconstruction, which we term now $x_{\text{prior}}$.
Motivated by the \fastmri challenge guidelines where \gls{ssim} is the major evaluation metric, we allow the fine-tuned reconstructions to deviate from the initial reconstruction $x_{\text{prior}}$ by a certain amount $\gamma_{\text{th}} \in [0, 1]$.
The influence of the regularizer is handled by the parameter $\gamma_{\text{prior}} > 0$.
The hyper-parameters were set to $\gamma_{\text{prior}}{=}1$ and $\gamma_{\text{th}}{=}0.8$ empirically and might be adapted for specific applications.

\subsection{Ensembling}
When training single models, we observe that the final reconstructions appear very similar at first glance.
We note only subtle differences, not only in different areas in the image, but also in terms of quantitative values, which might end in a great dissatisfaction that apparently different models lead to similar results.
This can be noted even when the same models are trained with different initializations, as the highly non-convex networks are trapped in different local minima.
As these models make subtle, different errors in the reconstruction, it becomes challenging to select a single \textit{best} model for a specific task.
To overcome these uncertainties, the reconstruction models trained under the same conditions are ensembled by simple averaging.
This increases the robustness not only quantitatively but also qualitatively~\cite{Goodfellow2016}, and was also used for ImageNet to yield top performances~\cite{Krizhevsky2012}.
In our work, we use model ensembling on all networks defined in a \sigmanet to balance the trade-off between quantitative scores and qualitative image appearance.

\section{Methods}

\subsection{Data Processing}
All our experiments were performed on the \fastmri knee dataset~\cite{Zbontar2018}.
Instead of using the \gls{rss} target of the \fastmri dataset, we re-defined the target as the sensitivity-weighted coil-combined image of the fully sampled data, which allowed us a valid comparison for any approach based on coil sensitivity maps.
We estimated two sets ($M{=}2$) of sensitivity maps according to soft SENSE~\cite{Uecker2014} to account for any field-of-view issues or other obstacles in the data.
The number of \glspl{acl} needed for sensitivity map estimation varied according to the acceleration factor and was set to 30 \glspl{acl} for $R{=}4$ and 15 \glspl{acl} for $R{=}8$ for the training and validation set.
Sensitivity maps for the test and challenge dataset were computed based on the provided number of low frequencies.
These numbers were motivated by examining the number of given low frequencies in the test and challenge dataset.
The data were normalized by a factor obtained from the low frequency scans by taking the median value of the 20\% largest magnitude values, to account for outliers.
Complex mean and complex pseudo co-variance normalization parameters for the \gls{dun} were also estimated from the low-frequency scans.

The \fastmri dataset contained patient volumes where each slice had a matrix size of $(N_\text{FE}\times N_\text{PE})$.
The \gls{pe} direction varied for the individual cases and the \gls{fe} direction was constant $N_\text{FE}{=}640$.
Including the $Q{=}15$ coil channels and estimated sensitivity maps, this required 2.8~TB for the train and validation set.
However, this resulted in a bottleneck for data reading itself.
To overcome this bottleneck and accelerate the training, we pre-processed the individual channels of the fully-sampled data in \gls{fe} direction and crop them to the central $N_\text{FE}{=}320$ while keeping the \gls{pe} the same, to not introduce new artifacts.
Furthermore, data were stored as float16, which in total reduced the memory to 750~GB.
Testing was performed on the original data.

As aforementioned in the section about supervised learning, we utilised the foreground information.
The foreground masks for 10 subjects were semi-automatically generated by first using graph-cut algorithms \cite{Rother2004} to obtain a rough segmentation, followed by manually applying standard image processing techniques including connected component analysis, hole-filling, binary mask dilation and erosion.
These images were treated as ground truth to train a \unet for foreground segmentation.
The trained network was applied to remainder of the data to obtain the reference foreground masks.
The underlying \unet architecture consisted of $n_f{=}32$ features with 4 downsampling levels. We note that perfect foreground masks are not necessary and one might also estimate foreground masks from the coil sensitivity maps which was not investigated in this work.
The estimated foreground masks will be provided upon request.

\subsection{Network Training}
We established a common training scheme for all networks, using RMSProp with learning rate $10^{-4}$ and batch size 1.
We pre-trained all network architectures on both contrasts PD and FS for the acceleration factors $R{=}4$ and $R{=}8$ simultaneously for 40 epochs and randomly selected slices from the individual training cases.
These networks were further fine-tuned for the individual contrasts and acceleration factors for 10 epochs using all available slices.
To overcome the huge \gls{gpu} memory consumption of the proposed networks, we further used a patch learning strategy~\cite{Schlemper2018b} where we randomly extracted patches of size 96 in \gls{fe} direction.

Similarly, networks with adversarial losses were also trained for each individual contrast and acceleration factor for 10 epochs based on the pre-trained models.
Adam optimizer with a learning rate of $10^{-4}$ was used to train the discriminator, while the settings for training the generator remained the same with the base model for a fair comparison.
Fine-tuning was performed for 50 iterations using the Adam optimizer with learning rate of $5\times10^{-5}$ at test time.

\subsection{Experimental Setup}
We performed two sets of experiments.
In the first set, we compareed the influence of different \gls{dc} layers.
For \glspl{sn}, we studied \gls{gd}, \gls{pg} and \gls{vs} for image reconstruction and additionally used the identity (ID) to compare to image enhancement, for $T{=}8$ cascades.
We refer to these networks as \sn{GD}{8}, \sn{PG}{8}, \sn{VS}{8} and \sn{ID}{8}, where the superscript denotes the number of cascades.
As a baseline, we trained a \gls{sn} for a single cascade $T{=}1$, denoted as \sn{ID}{1}, and a residual \unet.
For \glspl{pcn}, we performed experiments for \gls{pg}, denoted as \pcn{PG}{8}.
All experiments were performed for shared parameters and varying parameters over the cascades, where the latter case increased the model complexity by the number of cascades as shown in \cref{stab:stable1-num-params}.
Models trained with shared parameters are denoted by an additional $s$ in the superscript, \eg \sn{GD}{8s}.

In the second set of experiments, we studied the influence of \gls{lsgan}, semi-supervised fine-tuning and model ensembling on the qualitative and quantitative image quality.
These experiments were performed using \sn{GD}{8} as a base network.

Evaluation was performed on the \fastmri multicoil validation set.
For quantitative evaluation, we used the same evaluation procedure as in~\cite{Zbontar2018}, including \gls{nmse}, \gls{psnr} and \gls{ssim}.
Additionally, we studied the differences in \gls{dc} by plotting the dataterm argument $Ax - y$ and reporting
\begin{align*}
    \mathcal{D}\text{-NMSE}=\frac{1}{N_{\text{sl}}}\sum\limits_{n=1}^{N_{\text{sl}}}\frac{\norm[2]{A_nx_n - y_n}^2}{\norm[2]{y_n}^2}
\end{align*}
for a single case with $N_{\text{sl}}$ slices.

Statistical tests were performed for each metric result to ensure that the difference in model performances were significant.
For multi-model comparisons, we first performed Friedman test \cite{Friedman1937} to see if there was a significant difference in the result statistics.
Once the difference was detected, we performed one-versus-all one-way Wilcoxon signed-rank test~\cite{Wilcoxon1992} with Bonferroni correction to ensure if the results from a particular model significantly differed, \ie outperformed, the others.

\subsection{Reproducible Research}
Especially in the era of machine learning, it becomes more and more challenging to reproduce different approaches because they rely on their own framework and datasets.
With our work, we provide a unified framework for various regularization networks and \gls{dc} layers, allowing for a fair comparison between the different network architectures.
To support the reproducible research initiative~\cite{Stikov2019} and to verify the results of this manuscript, we make all Pytorch source code available at \url{https://github.com/khammernik/sigmanet}.
We built upon the \fastmri framework\footnote{\url{https://github.com/facebookresearch/fastMRI}} and provide not only improved data loaders, but also an interface for regularization and \gls{dc} layers, which can be easily extended.
The framework also includes data processing scripts to reduce data and to estimate both foreground masks and coil sensitivity maps.
To estimate the coil sensitivity maps, the BART toolbox~\cite{Uecker2015a} is required.
All coil sensitivity maps are stored in  \texttt{*.h5} along with other important variables.
A detailed description of these files is provided in Appendix A.
%

\section{results}

In the first set of experiments we compare the influence of different \gls{dc} layers on parallel \gls{mr} image reconstruction using \glspl{sn} and \glspl{pcn}, with shared and variable network parameters.
The quantitative results are depicted in \cref{tab:table1-dc}.
We observe that the proposed \gls{dun} regularization \sn{ID}{1} yields improved results compared to the \unet regularizer, depicted in the first two rows of the table.
Furthermore, we note that quantitative results can be pushed by training a separate \sn{ID}{8s}, which represents a deep residual \gls{nn} which has the same model complexity as \sn{ID}{1}.
\cref{tab:table1-dc} also shows that any type of reconstruction network outperforms all image enhancement networks, \ie residual U-net and \sn{ID}{*}, substantially.
For shared parameters, we note that reconstruction networks with \gls{pg} as \gls{dc} outperform \gls{gd} and \gls{vs} for both acceleration factors, while \gls{vs} outperforms \gls{gd} slightly only for acceleration factor $R{=}8$.
For variable parameters in the cascades, we observe that \gls{gd} as \gls{dc} performs superior than \gls{pg} and \gls{vs} in all cases.
Furthermore, we note that \glspl{pcn} yield inferior quantitative results than \glspl{sn}.
All results were statistically significant with $p \ll 10^{-5}$.

The qualitative results for different \gls{dc} layers is depicted in \cref{fig:figure3-img_dc_PDFS_R4} for a PDFS case and $R{=}4$.
We observe that the reconstruction networks appear not only sharper than image enhancement networks, but also depict the correct anatomy of the lateral meniscus, indicated by the green arrow.
There is no obvious difference between the different \gls{dc} layers \gls{gd}, \gls{pg} and \gls{vs} and also between \glspl{pcn} and \glspl{sn}.
The result for \sn{PG}{8s} appears slightly sharper and contains more noise than all other images.
Similar observations can be made for a PD case at acceleration factor $R{=}8$ as illustrated in \cref{fig:figure4-img_dc_PD_R8}, where the anatomy in the lateral aspect of the knee greatly deviates, especially for the \sn{ID}{8}.
These deviations in anatomy are depicted in the detailed view of \cref{fig:figure5-img_dc_detail}.
Additionally, we note for the PD case at acceleration $R{=}8$ that some bright intensities in the lateral meniscus area are not visible in neither image enhancement nor the image reconstruction networks, indicated by the green arrow.

In the second set of experiments, we compare the influence of different (semi-)supervised learning strategies which were conducted on the base network \sn{GD}{8}.
The quantitative results for this set of experiments are reported in \cref{tab:table2-loss}.
Qualitative results for a PDFS case is depicted in \cref{fig:figure6-img_loss_PDFS_R8} along with a detailed view of areas of the medial and lateral aspect of the knee and the interchondylar notch.
Quantitative values indicate that the base network \sn{GD}{8} achieves the best scores, followed by the ensembled \sigmanetarch.
The quantitative values for \sn{GD-FT}{8} and \sn{GD-GAN}{8} drop substantially for all cases.
In the qualitative results, we observe that the base network \sn{GD}{8} is most blurred.
The texture is improved for \sn{GD-FT}{8}, while \sn{GD-GAN}{8} appears most textured.
The ensembled \sigmanetarch combines the properties of different \glspl{sn} and appears slightly more textured than the base network \sn{GD}{8} with minimally decreased quantitative scores.

To support our findings in the importance of \gls{dc}, we study the \gls{dc} qualitatively and quantitatively for a single PDFS case at acceleration $R{=}4$ for \sn{ID}{8}, \sn{GD}{8}, \sn{GD-FT}{8} and \sn{GD-GAN}{8} in \cref{sfig:sfigure3-datanorm}.
We observe the same behaviour: The lowest \gls{ssim} score is achieved for \sn{ID}{8} and the highest \gls{ssim} score for \sn{GD}{8}.
Scores for \sn{GD-FT}{8} and \sn{GD-GAN}{8} dropped again compared to \sn{GD}{8}.
The dataterm measure \dnmse indicates best \gls{dc} for \sn{GD-FT}{8}, followed by \sn{GD-GAN}{8} and \sn{GD}{8}.
\gls{dc} is not maintained in \sn{ID}{8}, also supported by the \kspace plot $Ax-y$ of a depicted coil.
Interestingly, the \dnmse is lower for \sn{GD-GAN}{8} than \sn{GD}{8}, however, we observe a bright spot in the center of \kspace for \sn{GD-GAN}{8}.
We note that the improved textures and the drop in \gls{ssim} of \sn{GD-FT}{8} from \sn{GD}{8} is consistent with \cref{eq:semisupervised}, hence, expected.

\section{Discussion}

To the best of our knowledge, this is the first large-scale comparison of \sota approaches based on a systematic evaluation of different model configurations with the same model complexity.
Our first set of experiments is concerned with comparing the effect of different \gls{dc} layers.
In particular, \sn{ID}{} represents any image enhancement networks like~\cite{Jin2017,Lee2017a,Zbontar2018}, \sn{GD}{8} represents \cite{Hammernik2018,Schlemper2018a}, \sn{GD-GAN}{8} represents \cite{Mardani2018}, \sn{GD-PG}{8s} represents MoDL \cite{Aggarwal2018}, \sn{GD-VS}{8} represents \gls{vs}-net \cite{Duan2019} and \pcn{PG}{8} represents \cite{Schlemper2019,Wang2019b}.
Note that the difference to the original approaches is that we have employed the same base architecture for the regularization network, realized by \glspl{dun}, and a fixed number of iterations ($T{=}8$).
This systematic evaluation allows us to compare the different approaches in a fairer and more informative way, while re-implementation of the original architectures would yield in different model complexities, \ie number of parameters, regarding the regularization networks.
From the results we have observed that while \gls{gd} outperforms other \gls{dc} types in many scenarios quantitatively, perceptually most models result in similar performance for the underlying experimental setup.
Note that we have not compared with other extensions of unrolled approaches~\cite{Qin2017} but we believe that these extensions can mutually benefit all models, in exchange of higher memory consumption.

Regarding the training setting, we have used the same combination that worked robust across all scenarios to enable a fair comparison between different models in a reproducible way.
We have used \glspl{dun} (\sn{ID}{}) as the base architecture, which indeed outperforms \unet for 70\% of memory utilization.
This ensures sufficient expressibility of the model, yet the efficient memory usage is crucial to enable a certain number of cascades for the experiments.
It is important to note that we expect further improved results by elaborating other network architectures, fine-tuning the learning rate with, \eg scheduling, and/or curriculum learning.
Again, these strategies cannot be tested extensively as they may be model specific strategies that are based on empirical experience only.
This is out of scope for this work.
However we do not expect that the general trend will change according to our observations.


Besides different training procedures and \gls{dc} layers,  we also compare two different types of network architectures, \glspl{sn} and \glspl{pcn}.
For \glspl{sn} we provide explicit coil sensitivity maps and the coil combination is performed by the forward operator $A$, whereas \glspl{pcn} have single-coil \gls{dc} and the weighted coil combination is performed implicitly by the regularizing \gls{dun}.
Both proposed architectures have approximately the same model complexity.
The result indicate that both \glspl{sn} and \glspl{pcn} outperform image enhancement networks significantly.
While large differences between \glspl{sn} and \glspl{pcn} are hard to spot in general, the quantitative results that training of \glspl{sn} is facilitated by providing some additional prior knowledge through the forward operator $A$.
The advantage of \glspl{pcn} are that they do not require coil sensitivity maps.


One of the main aims of this work is to examine the benefits of image reconstruction approaches compared to image enhancement approaches.
We have found that image reconstruction approaches significantly outperform image enhancement approaches $(p \ll 10^{-5})$.
However, we have observed that even without \gls{dc}, high quantitative results can be achieved.
Nevertheless, while the reconstruction results appear correct qualitatively, the anatomy appears completely different in certain areas compared to the reference, as illustrated in \cref{fig:figure5-img_dc_detail}, which might result in mis-diagnosis.
The enhancement approaches overfit severely, even if the overall result seems consistent.
The reconstruction approaches have better \gls{dc}, hence, anatomical consistency. 
However, for $R{=}8$, we see a loss in content for both models, highlighted by the green arrow in \cref{fig:figure5-img_dc_detail}, which may indicate the limit of acceleration in static 2D \gls{mri} reconstruction.

Our systematic experiments also study the influence of shared and variable parameters across the individual cascades.
While variable parameters increase the model complexity by the number of cascades $T$, the same set of parameters is applied to the individual cascades in the shared setting (see \cref{stab:stable1-num-params}).
Our results support the findings of Aggarwal \etal~\cite{Aggarwal2018}, who also used shared parameters but a different regularization network.
However, we have found that the gap between \sn{GD}{8s} and \sn{PG}{8s} is smaller compared the findings in~\cite{Aggarwal2018} which might be due to a different network training procedure, imperfect validation set, different data set and most importantly, the high sensitivity of the trainable parameter $\lambda$.
For shared parameters, we also observe that the training of \gls{vs} is very sensitive to parameter selection and the training parameters, resulting in decreased values for acceleration factor $R{=}4$ and improved results for acceleration factor $R{=}8$ compared to \gls{gd}.
The ambiguities in splitting methods, both \gls{pg} and \gls{vs}, are more obvious when examining the results for variable parameters.
We clearly observe a decrease in quantitative values and \gls{dc} with \gls{gd} outperforms all other methods in the case of a variable parameter setting.
We also observe instabilities during training the different architectures and a huge influence of the parameter $\lambda$.
These findings also suggest that we have to carefully setup our training procedure and either keep $\lambda$ and other hyper-parameters fixed or optimize with a different learning rate than the convolution kernels.
We also find that the quantitative results further improve by having a greater model complexity, however, this also rises the question if it is really required to rise this complexity by such a large margin for a small improvement in quantitative values, or if more focus should be given on a careful evaluation of a specific application in future work.


The second set of experiments investigates the effect of adversarial loss, semi-supervised fine-tuning and model ensembling.
%
%
We have observed that models trained with an adversarial loss can reconstruct images with visually better details and textures, though with a compromise of decreased quantitative values, which is consistent with preliminary findings \cite{Shitrit2017,Seitzer2018,Quan2017,Yang2017,Liu2017,Mardani2018,Hammernik2018a}.
In addition, we note that combination of \gls{dc} layers and adversarial loss is complementary and improved sharpness of the results with minimal risk of hallucination.
Therefore, we have not made direct comparisons with methods such as DAGAN \cite{Yang2017} and CycleGAN \cite{Liu2017}, as we have observed that \sn{GD}{8} served as a superior base network than U-net.
Nevertheless, we have not utilized other components proposed in literature such as perceptual loss, $k$-space loss and cycle consistency, but we expect that the addition of these elements can provide further improvements.
Similarly, we use \gls{lsgan}  as we find it to be stable, but other types of \glspl{gan}, \eg Wasserstein \gls{gan} \cite{Mardani2018}, can be considered, even though extensive comparisons are beyond the scope of this work.

On the other hand, the semi-supervised approach fine-tunes each test image by considering the specific features and characteristics of the new unseen \kspace data, while exploiting the prior knowledge learned from the supervised learning stage.
Results show that the semi-supervised fine-tuning can recover more textures compared to the models trained with the supervised base loss alone, with more details that are specific to the test sample. We note that the fine-tuning process results in lower quantitative values.
While this may appear counter intuitive, it is consistent with \cref{eq:semisupervised} that data-specific information is recovered even though it may not directly optimize the metric of interest.
%
%
Our findings support that the balance between the perceptual quality and data fidelity shall be considered.
This can be achieved via varying the trade-off parameters between different losses to find an optimal balance between them.
However, the aim of this work is also to investigate the effects of these losses, thus we do not further investigate an optimal tuning of these parameters.

In addition, we have also proposed to ensemble models with different network structures and different training losses, resulting in the \sigmanet.
Model ensembling can be seen as a tool to remove random errors made by individual reconstructions.
If the single model performances are on similar lines, this might push the quantitative values while exploiting the advantages from different models.
It can be seen from our results that model ensembling achieves a balancing effect between different models, producing less smooth images compared to \sn{GD}{8}, with relatively close quantitative scores.
The current performance of model ensembling is limited by the diversities of available models, as most of them adopt the same network architecture and produce similar results.
If more models with various similar architectures are included, we could expect a even better performance of model ensembling.


In this work, we have studied many variants for learning parallel \gls{mr} image reconstruction systematically.
In general, deep learning approaches have shown robust, near perfect result for $R{=}4$ and also impressive results for $R{=}8$.
However, while higher quantitative scores correlate with improved image quality in general, they simply cannot guarantee if local structures are accurately depicted, because the provided measures are not sensitive to this. 
This necessitates careful side-by-side comparisons such that the radiologist agrees on local, anatomical consistency.
Moreover, metrics which are more sensitive to local consistencies might be a key ingredient to assess the algorithms quantitatively in the future.

Another concern is that the current limit of machine learning based algorithms for inverse problem is unclear.
While this limit seems to be much further away than the current theoretical limit of \gls{cs}, it is unclear how far the acceleration factor should be pushed.
An acceleration factor of $R{=}8$ might lead to mis-diagnosis, as our networks cannot recover the bright intensities in the lateral meniscus, as illustrated in \cref{fig:figure5-img_dc_detail}.
Hence, the acceleration limit of \gls{mr} data acquisition highly depends on the specific purpose and more application-specific approaches might be studied in the future, where higher acceleration can be considered~\cite{Caballero2014a,Schlemper2018}.
%

\section{Conclusion}
In this work, we systematically studied different configurations of \sota network designs for parallel \gls{mr} image reconstruction in image domain, including \glspl{dun}, \unet and different \gls{dc} terms.
To further enhance the image quality, we investigate adversarial losses and self-supervised fine-tuning.
The final model ensembling to a \sigmanet takes advantages of the strengths and weaknesses of the included network architectures and allows for both balanced quantitative scores and enhanced image quality.
The proposed \sigmanet of our team \holykspace is among the top three performing methods on the \fastmri public and challenge leaderboard for multicoil reconstruction~\cite{Schlemper2019a}.
In this work, we used a sensitivity-weighted coil-combination as reference, different from the \gls{rss} reference used in fastMRI challenge.
This enabled us to minimize the effect of noise-induced bias and perform fairer comparisons of multicoil models.

This work provides general open-source tools for learning-based \gls{mri} reconstruction in image domain and provides an overview of various proposed algorithms in literature, along with their challenges in, \eg network training.
We emphasize that each model definitely has scope for further improvement based on careful hyper-parameter tuning, including variations in initialization, advanced (pre-)training schemes, different loss functions and learning rate scheduling.
However, this is out-of-scope for this work as we highlight that all models were trained systematically in a common setting.

Up to now, we have only scratched the surface of deep learning for \gls{mr} image reconstruction, we are yet to understand their generalization ability to, \eg variations in anatomy, noise, contrast, as well as finding the failing mode \cite{Antun2019}.
The \fastmri dataset is a big step towards reproducible \gls{mr} image reconstruction research for studying the model performances, nevertheless, it raised new challenges.
In particular, radiologists' perspective and quantitative scores did not align for local characteristics of the images, suggesting that one should direct the research in both careful network setup according to the underlying target application and finding new evaluation schemes to judge, which algorithms fulfill the target application's requirements.
	
\vspace*{1\baselineskip} \section{Acknowledgements}
The work was funded in part by the EPSRC Programme Grant (EP/P001009/1) and by the Intramural Research Programs of the National Institutes of Health Clinical Center (1Z01 CL040004).

\section{Appendix A}\label{sec:smaps}

\subsection{Description of Sensitivity Map Files}
We followed the file descriptions of~\cite{Zbontar2018} and generate HDF5 files with the same filenames as the original \fastmri dataset.
We computed an extended set of coil sensitivity maps ($M{=}2$) according to Uecker \etal~\cite{Uecker2014}.
The sensitivity maps were computed on the \glspl{acl}, defined by \texttt{num\_low\_frequency}. For training and validation data \texttt{num\_low\_frequency} was set to 30, used for acceleration factor $R\redequal4$, and 15, for $R\redequal8$, respectively. For testing and challenge data, the \texttt{num\_low\_frequency} was defined in the HDF5 raw data files.
Following data are stored in a single HDF5 files:

\paragraph{reference\_acl\{\texttt{num\_low\_frequency}\}} [$N_{\text{sl}}$, $M$, $N_{\text{FE}}$, $N_{\text{PE}}$]. Coil-combined reference reconstruction. This dataset is only available for training and validation.
\paragraph{smaps\_acl\{\texttt{num\_low\_frequency}\}} [$N_{\text{sl}}$, $Q$, $M$, $N_{\text{FE}}$, $N_{\text{PE}}$]. Estimated coil sensitivity maps.
\paragraph{ismrmrd\_header} XML header copied from the original HDF5 raw data file.
\paragraph{norm\_lfimg\_max\_acl\{\texttt{num\_low\_frequency}\}} Attribute. Maximum intensity of the [$N_{\text{sl}}$,~320,~320] coil-combined low frequency reconstruction.
\paragraph{norm\_lfimg\_mean\_acl\{\texttt{num\_low\_frequency}\}} Attribute. Mean complex intensity of the [$N_{\text{sl}}$,~320,~320] coil-combined low frequency reconstruction for the first sensitivity map.
\paragraph{norm\_lfimg\_cov\_acl\{\texttt{num\_low\_frequency}\}} Attribute. Complex pseudo-covariance of the [$N_{\text{sl}}$,~320,~320] coil-combined low frequency reconstruction for the first sensitivity map.
\paragraph{reference\_max\_acl\{\texttt{num\_low\_frequency}\}} Maximum magnitude value of the [$N_{\text{sl}}$,~320,~320] coil-combined fully sampled reconstruction. This attribute is only available for training and validation.
\paragraph{rss\_max} Maximum value of the [$N_{\text{sl}}$,~320,~320] \gls{rss} reconstruction. This attribute is only available for training and validation.

\clearpage \section{References}
\bibliographystyle{mrm}
\bibliography{main}

\begin{thebibliography}{10}

\bibitem{Sodickson1997}
Sodickson DK, Manning WJ.
\newblock {Simultaneous Acquisition of Spatial Harmonics (SMASH): Fast Imaging
  With Radiofrequency Coil Arrays}.
\newblock Magnetic Resonance in Medicine 1997;\hspace{0pt}38(4):591--603.

\bibitem{Pruessmann2001}
Pruessmann KP, Weiger M, Boernert P, Boesiger P.
\newblock {Advances in Sensitivity Encoding With Arbitrary k-Space
  Trajectories}.
\newblock Magnetic Resonance in Medicine 2001;\hspace{0pt}46(4):638--651.

\bibitem{Griswold2002}
Griswold MA, Jakob PM, Heidemann RM, Nittka M, Jellus V, Wang J, Kiefer B,
  Haase A.
\newblock {Generalized Autocalibrating Partially Parallel Acquisitions
  (GRAPPA)}.
\newblock Magnetic Resonance in Medicine 2002;\hspace{0pt}47(6):1202--1210.

\bibitem{Uecker2014}
Uecker M, Lai P, Murphy MJ, Virtue P, Elad M, Pauly JM, Vasanawala SS, Lustig
  M.
\newblock {ESPIRiT--An Eigenvalue Approach to Autocalibrating Parallel MRI:
  Where SENSE Meets GRAPPA}.
\newblock Magnetic Resonance in Medicine 2014;\hspace{0pt}71(3):990--1001.

\bibitem{Lustig2006}
Lustig M, Santos JM, Donoho DL, Pauly JM.
\newblock {k-t SPARSE: High frame rate dynamic MRI exploiting spatio-temporal
  sparsity}.
\newblock In Proceedings of the International Society of Magnetic Resonance in
  Medicine. 2006;\hspace{0pt} 2420.

\bibitem{Liang2009}
Liang D, Liu B, Wang J, Ying L.
\newblock {Accelerating SENSE Using Compressed Sensing}.
\newblock Magnetic Resonance in Medicine 2009;\hspace{0pt}62(6):1574--1584.

\bibitem{Murphy2012}
Murphy M, Alley M, Demmel J, Keutzer K, Vasanawala S, Lustig M.
\newblock {F}ast $\ell_1$-{SPIR}i{T} {C}ompressed {S}ensing {P}arallel
  {I}maging {MRI}: {S}calable {P}arallel {I}mplementation and {C}linically
  {F}easible {R}untime.
\newblock IEEE Transactions on Medical Imaging
  2012;\hspace{0pt}31(6):1250--1262.

\bibitem{Shin2014}
Shin PJ, Larson PEZ, Ohliger MA, Elad M, Pauly JM, Vigneron DB, Lustig M.
\newblock {Calibrationless Parallel Imaging Reconstruction Based on Structured
  Low-Rank Matrix Completion}.
\newblock Magnetic Resonance in Medicine 2014;\hspace{0pt}72(4):959--970.

\bibitem{Haldar2013}
Haldar JP.
\newblock Low-rank modeling of local $k$-space neighborhoods (loraks) for
  constrained mri.
\newblock IEEE Transactions on Medical Imaging 2013;\hspace{0pt}33(3):668--681.

\bibitem{Haldar2016}
Haldar JP, Zhuo J.
\newblock {P-LORAKS: Low-Rank Modeling of Local k-Space Neighborhoods With
  Rarallel Imaging Data}.
\newblock Magnetic Resonance in Medicine 2016;\hspace{0pt}75(4):1499--1514.

\bibitem{Jin2016}
Jin KH, Lee D, Ye JC.
\newblock {A General Framework for Compressed Sensing and Parallel MRI Using
  Annihilating Filter Based Low-Rank Hankel Matrix}.
\newblock IEEE Transactions on Computational Imaging
  2016;\hspace{0pt}2(4):480--495.

\bibitem{Block2007}
Block KT, Uecker M, Frahm J.
\newblock {Undersampled Radial MRI with Multiple Coils. Iterative Image
  Reconstruction using a Total Variation Constraint.}
\newblock Magnetic Resonance in Medicine 2007;\hspace{0pt}57(6):1086--1098.

\bibitem{Knoll2012}
Knoll F, Clason C, Bredies K, Uecker M, Stollberger R.
\newblock {Parallel Imaging with Nonlinear Reconstruction using Variational
  Penalties}.
\newblock Magnetic Resonance in Medicine 2012;\hspace{0pt}67(1):34--41.

\bibitem{Ravishankar2012}
Ravishankar S, Bresler Y.
\newblock {Learning Sparsifying Transforms for Image Processing}.
\newblock IEEE International Conference on Image Processing
  2012;\hspace{0pt}61(5):681--684.

\bibitem{Caballero2014}
Caballero J, Price AN, Rueckert D, Hajnal JV.
\newblock {Dictionary Learning and Time Sparsity for Dynamic MR Data
  Reconstruction}.
\newblock IEEE Transactions on Medical Imaging 2014;\hspace{0pt}33(4):979--994.

\bibitem{Hammernik2018}
Hammernik K, Klatzer T, Kobler E, Recht MP, Sodickson DK, Pock T, Knoll F.
\newblock {Learning a Variational Network for Reconstruction of Accelerated MRI
  Data}.
\newblock Magnetic Resonance in Medicine 2018;\hspace{0pt}79(6):3055--3071.

\bibitem{Wang2016ab}
Wang S, Su Z, Ying L, Peng X, Zhu S, Liang F, Feng D, Liang D.
\newblock {Accelerating Magnetic Resonance Imaging Via Deep Learning}.
\newblock In IEEE International Symposium on Biomedical Imaging (ISBI).
  2016;\hspace{0pt} 514--517.

\bibitem{Han2018}
Han Y, Yoo J, Kim HH, Shin HJ, Sung K, Ye JC.
\newblock {Deep learning with domain adaptation for accelerated
  projection-reconstruction MR}.
\newblock Magnetic Resonance in Medicine 2018;\hspace{0pt}80(3):1189--1205.

\bibitem{Zhu2018}
Zhu B, Liu JZ, Cauley SF, Rosen BR, Rosen MS.
\newblock {Image reconstruction by domain-transform manifold learning}.
\newblock Nature 2018;\hspace{0pt}555(7697):487--492.

\bibitem{Schlemper2018a}
Schlemper J, Yang G, Ferreira P, Scott A, McGill LA, Khalique Z, Gorodezky M,
  Roehl M, Keegan J, Pennell D, Firmin D, Rueckert D.
\newblock {Stochastic Deep Compressive Sensing for the Reconstruction of
  Diffusion Tensor Cardiac MRI}.
\newblock arXiv preprint arXiv:180512064 2018;\hspace{0pt}.

\bibitem{Yang2016}
Yang Y, Sun J, Li H, Xu Z.
\newblock {ADMM-Net: A Deep Learning Approach for Compressive Sensing MRI}.
\newblock In Advances in Neural Information Processing Systems.
  2017;\hspace{0pt} 10--18.

\bibitem{Zhao2015}
Zhao H, Gallo O, Frosio I, Kautz J.
\newblock {Loss Functions for Image Restoration with Neural Networks}.
\newblock IEEE Transactions on Computational Imaging 2016;\hspace{0pt}3(1):47
  -- 57.

\bibitem{Hammernik2017f}
Hammernik K, Knoll F, Sodickson D, Pock T.
\newblock {L2 or Not L2: Impact of Loss Function Design for Deep Learning MRI
  Reconstruction}.
\newblock In Proceedings of the International Society of Magnetic Resonance in
  Medicine. 2017;\hspace{0pt} 687.

\bibitem{Yang2017}
Yang G, Yu S, Dong H, Slabaugh G, Dragotti PL, Ye X, Liu F, Arridge S, Keegan
  J, Guo Y, Firmin D.
\newblock {DAGAN: Deep De-Aliasing Generative Adversarial Networks for Fast
  Compressed Sensing MRI Reconstruction}.
\newblock IEEE Transactions on Medical Imaging
  2018;\hspace{0pt}37(6):1310--1321.

\bibitem{Liu2017}
Liu F, Samsonov A.
\newblock {Data-Cycle-Consistent Adversarial Networks for High-Quality
  Reconstruction of Undersampled MRI Data}.
\newblock In ISMRM Workshop on Machine Learning. 2017;\hspace{0pt} .

\bibitem{Quan2017}
Quan TM, Nguyen-Duc T, Jeong WK.
\newblock {Compressed Sensing MRI Reconstruction using a Generative Adversarial
  Network with a Cyclic Loss}.
\newblock IEEE Transactions on Medical Imaging
  2018;\hspace{0pt}37(6):1488--1497.

\bibitem{Seitzer2018}
Seitzer M, Yang G, Schlemper J, Oktay O, W{\"{u}}rfl T, Christlein V, Wong T,
  Mohiaddin R, Firmin D, Keegan J, Rueckert D, Maier A.
\newblock {Adversarial and perceptual refinement for compressed sensing MRI
  reconstruction}.
\newblock In Lecture Notes in Computer Science (including subseries Lecture
  Notes in Artificial Intelligence and Lecture Notes in Bioinformatics), volume
  11070 LNCS. 2018;\hspace{0pt} 232--240.

\bibitem{Mardani2018}
Mardani M, Gong E, Cheng JY, Vasanawala SS, Zaharchuk G, Xing L, Pauly JM.
\newblock {Deep Generative Adversarial Neural Networks for Compressive Sensing
  (GANCS) MRI}.
\newblock IEEE Transactions on Medical Imaging 2019;\hspace{0pt}38(1):167--179.

\bibitem{Aggarwal2018}
Aggarwal HK, Mani MP, Jacob M.
\newblock {MoDL: Model Based Deep Learning Architecture for Inverse Problems}.
\newblock IEEE Transactions on Medical Imaging 2019;\hspace{0pt}38(2):394--405.

\bibitem{Cheng2018}
Cheng JY, Mardani M, Alley MT, Pauly JM, Vasanawala SS.
\newblock {DeepSPIRiT: Generalized Parallel Imaging Using Deep Convolutional
  Neural Networks}.
\newblock In Proceedings of the International Society of Magnetic Resonance in
  Medicine. 2018;\hspace{0pt} 570.

\bibitem{Kwon2017a}
Kwon K, Kim D, Park H.
\newblock {A parallel MR imaging method using multilayer perceptron:}.
\newblock Medical Physics 2017;\hspace{0pt}44(12):6209--6224.

\bibitem{Han2018a}
Han Y, Ye JC.
\newblock {k-Space Deep Learning for Accelerated MRI}.
\newblock arXiv preprint arXiv:180503779 2018;\hspace{0pt}1--11.

\bibitem{Eo2018}
Eo T, Jun Y, Kim T, Jang J, Lee HJ, Hwang D.
\newblock {KIKI-net: Cross-domain convolutional neural networks for
  reconstructing undersampled magnetic resonance images}.
\newblock Magnetic Resonance in Medicine 2018;\hspace{0pt}80(5):2188--2201.

\bibitem{Souza2018a}
Souza R, Lebel RM, Frayne R, Ca R.
\newblock {A Hybrid, Dual Domain, Cascade of Convolutional Neural Networks for
  Magnetic Resonance Image Reconstruction}.
\newblock Proceedings of Machine Learning Research
  2019;\hspace{0pt}102:437--446.

\bibitem{Akcakaya2019}
Ak{\c{c}}akaya M, Moeller S, Weing{\"{a}}rtner S, U{\u{g}}urbil K.
\newblock {Scan-specific robust artificial-neural-networks for k-space
  interpolation (RAKI) reconstruction: Database-free deep learning for fast
  imaging}.
\newblock Magnetic Resonance in Medicine 2019;\hspace{0pt}81(1):439--453.

\bibitem{Virtue2017}
Virtue P, Stella XY, Lustig M.
\newblock {Better than Real: Complex-valued Neural Nets for MRI
  Fingerprinting}.
\newblock In IEEE International Conference on Image Processing. IEEE,
  2017;\hspace{0pt} 3953--3957.

\bibitem{Wang2019a}
Wang S, Cheng H, Ying L, Xiao T, Ke Z, Liu X, Zheng H, Liang D.
\newblock {DeepcomplexMRI: Exploiting deep residual network for fast parallel
  MR imaging with complex convolution}.
\newblock arXiv preprint arXiv:190604359 2019;\hspace{0pt}.

\bibitem{Sun2019}
Sun L, Fan Z, Fu X, Huang Y, Ding X, Paisley J.
\newblock {A Deep Information Sharing Network for Multi-Contrast Compressed
  Sensing MRI Reconstruction}.
\newblock IEEE Transactions on Image Processing
  2019;\hspace{0pt}28(12):6141--6153.

\bibitem{Polak2019}
Polak D, Cauley S, Bilgic B, Gong E, Bachert P, Adalsteinsson E, Setsompop K.
\newblock {Joint multi-contrast Variational Network reconstruction (jVN) with
  application to rapid 2D and 3D imaging}.
\newblock arXiv preprint arXiv:191003273 2019;\hspace{0pt}.

\bibitem{Qin2017}
Qin C, Schlemper J, Caballero J, Price AN, Hajnal JV, Rueckert D.
\newblock {Convolutional Recurrent Neural Networks for Dynamic MR Image
  Reconstruction}.
\newblock IEEE Transactions on Medical Imaging 2019;\hspace{0pt}38(1):280--290.

\bibitem{Schlemper2018b}
Schlemper J, Caballero J, Hajnal JV, Price AN, Rueckert D.
\newblock {A Deep Cascade of Convolutional Neural Networks for Dynamic MR Image
  Reconstruction}.
\newblock IEEE Transactions on Medical Imaging 2018;\hspace{0pt}37(2):491--503.

\bibitem{Qin2019}
Qin C, Schlemper J, Duan J, Seegoolam G, Price A, Hajnal J, Rueckert D.
\newblock {k-t NEXT: Dynamic MR Image Reconstruction Exploiting Spatio-Temporal
  Correlations}.
\newblock In International Conference on Medical Image Computing and
  Computer-Assisted Intervention. Springer, 2019;\hspace{0pt} 505--513.

\bibitem{Hauptmann2018a}
Hauptmann A, Arridge S, Lucka F, Muthurangu V, Steeden JA.
\newblock {Real-time Cardiovascular MR with Spatio-temporal Artifact
  Suppression Using Deep Learning-Proof of Concept in Congenital Heart
  Disease}.
\newblock Magnetic Resonance in Medicine 2019;\hspace{0pt}81(2):1143--1156.

\bibitem{Jin2019}
Jin KH, Gupta H, Yerly J, Stuber M, Unser M.
\newblock {Time-Dependent Deep Image Prior for Dynamic MRI}.
\newblock arXiv preprint arXiv:191001684 2019;\hspace{0pt}.

\bibitem{Kofler2019}
Kofler A, Dewey M, Schaeffter T, Wald C, Kolbitsch C.
\newblock {Spatio-Temporal Deep Learning-Based Undersampling Artefact Reduction
  for 2D Radial Cine MRI with Limited Data}.
\newblock arXiv preprint arXiv:190401574 2019;\hspace{0pt}.

\bibitem{Cohen2018}
Cohen O, Zhu B, Rosen MS.
\newblock {MR fingerprinting Deep RecOnstruction NEtwork (DRONE)}.
\newblock Magnetic Resonance in Medicine 2018;\hspace{0pt}80(3):885--894.

\bibitem{Knoll2019}
Knoll F, Hammernik K, Zhang C, Moeller S, Pock T, Sodickson DK, Akcakaya M.
\newblock {Deep Learning Methods for Parallel Magnetic Resonance Image
  Reconstruction}.
\newblock arXiv preprint arXiv:190401112 2019;\hspace{0pt}.

\bibitem{Hammernik2018b}
Hammernik K, Knoll F.
\newblock {Machine Learning for Image Reconstruction}.
\newblock In D~Rueckert, G~Fichtinger, SK~Zhou, eds., Handbook of Medical Image
  Computing and Computer Assisted Intervention. Elsevier, 2018;\hspace{0pt}
  25--64.

\bibitem{Liang2019}
Liang D, Cheng J, Ke Z, Ying L.
\newblock {Deep MRI Reconstruction: Unrolled Optimization Algorithms Meet
  Neural Networks}.
\newblock arXiv preprint arXiv:190711711 2019;\hspace{0pt}.

\bibitem{Lundervold2019}
Lundervold AS, Lundervold A.
\newblock {An overview of deep learning in medical imaging focusing on MRI}.
\newblock Zeitschrift f{\"{u}}r Medizinische Physik
  2019;\hspace{0pt}29(2):102--127.

\bibitem{Zbontar2018}
Zbontar J, Knoll F, Sriram A, Muckley MJ, Bruno M, Defazio A, Parente M, Geras
  KJ, Katsnelson J, Chandarana H, Zhang Z, Drozdzal M, Romero A, Rabbat M,
  Vincent P, Pinkerton J, Wang D, Yakubova N, Owens E, Zitnick CL, Recht MP,
  Sodickson DK, Lui YW.
\newblock {fastMRI: An Open Dataset and Benchmarks for Accelerated MRI}.
\newblock arXiv preprint arXiv:181108839 2018;\hspace{0pt}.

\bibitem{Putzky2019}
Putzky P, Welling M.
\newblock {Invert to Learn to Invert}.
\newblock In Advances in Neural Information Processing Systems.
  2019;\hspace{0pt} 444--454.

\bibitem{Bahadir2019}
Bahadir CD, Dalca AV, Sabuncu MR.
\newblock {Adaptive Compressed Sensing MRI with Unsupervised Learning}.
\newblock arXiv preprint arXiv:190711374 2019;\hspace{0pt}.

\bibitem{Sriram2019}
Sriram A, Zbontar J, Murrell T, Zitnick CL, Defazio A, Sodickson DK.
\newblock {GrappaNet: Combining Parallel Imaging with Deep Learning for
  Multi-Coil MRI Reconstruction}.
\newblock arXiv preprint arXiv:191012325 2019;\hspace{0pt}.

\bibitem{Wang2019b}
Wang P, Chen EZ, Chen T, Patel VM, Sun S.
\newblock {Pyramid Convolutional RNN for MRI Reconstruction}.
\newblock arXiv preprint arXiv:191200543 2019;\hspace{0pt}.

\bibitem{Roemer1990}
Roemer PB, Edelstein WA, Hayes CE, Souza SP, Mueller OM.
\newblock {The NMR Phased Array}.
\newblock Magnetic Resonance in Medicine 1990;\hspace{0pt}16(2):192--225.

\bibitem{Robson2008}
Robson PM, Grant AK, Madhuranthakam AJ, Lattanzi R, Sodickson DK, McKenzie CA.
\newblock {Comprehensive Quantification of Signal-to-Noise Ratio and g-Factor
  for Image-based and k-Space-based Parallel Imaging Reconstructions}.
\newblock Magnetic Resonance in Medicine 2008;\hspace{0pt}60(4):895--907.

\bibitem{Jin2017}
Jin KH, McCann MT, Froustey E, Unser M.
\newblock {Deep Convolutional Neural Network for Inverse Problems in Imaging}.
\newblock IEEE Transactions on Image Processing
  2017;\hspace{0pt}26(9):4509--4522.

\bibitem{Lee2017a}
Lee D, Yoo J, Ye JC.
\newblock {Deep residual learning for compressed sensing MRI}.
\newblock In IEEE International Symposium on Biomedical Imaging.
  2017;\hspace{0pt} 15--18.

\bibitem{Schlemper2019}
Schlemper J, Duan J, Ouyang C, Qin C, Caballero J, Hajnal JV, Rueckert D.
\newblock {Data Consistency Networks for (Calibration-less) Accelerated
  Parallel MR Image Reconstruction}.
\newblock In Proceedings of the International Society of Magnetic Resonance in
  Medicine. 2019;\hspace{0pt} 4664.

\bibitem{Duan2019}
Duan J, Schlemper J, Qin C, Ouyang C, Bai W, Biffi C, Bello G, Statton B,
  O’Regan DP, Rueckert D.
\newblock {VS-Net: Variable Splitting Network for Accelerated Parallel MRI
  Reconstruction}.
\newblock In International Conference on Medical Image Computing and
  Computer-Assisted Intervention. 2019;\hspace{0pt} 713--722.

\bibitem{Schlemper2019a}
Schlemper J, Qin C, Duan J, Summers RM, Hammernik K.
\newblock $\Sigma$-net: {Ensembled Iterative Deep Neural Networks for
  Accelerated Parallel MR Image Reconstruction}.
\newblock arXiv preprint arXiv:191205480 2019;\hspace{0pt}.

\bibitem{Ronneberger2015}
Ronneberger O, Fischer P, Brox T.
\newblock {U-Net: Convolutional Networks for Biomedical Image Segmentation}.
\newblock In International Conference on Medical Image Computing and Computer
  Assisted Intervention. 2015;\hspace{0pt} 234--241.

\bibitem{Yu2019}
Yu S, Park B, Jeong J.
\newblock {Deep Iterative Down-Up CNN for Image Denoising}.
\newblock In Proceedings of the IEEE Conference on Computer Vision and Pattern
  Recognition Workshops. 2019;\hspace{0pt} .

\bibitem{Abdelhamed2019}
Abdelhamed A, Timofte R, Brown MS.
\newblock {NTIRE 2019 Challenge on Real Image Denoising: Methods and Results}.
\newblock In Proceedings of the IEEE Conference on Computer Vision and Pattern
  Recognition Workshops. 2019;\hspace{0pt} .

\bibitem{Shi2016a}
Shi W, Caballero J, Huszar F, Totz J, Aitken AP, Bishop R, Rueckert D, Wang Z.
\newblock {Real-Time Single Image and Video Super-Resolution Using an Efficient
  Sub-Pixel Convolutional Neural Network}.
\newblock In Proceedings of the IEEE Computer Society Conference on Computer
  Vision and Pattern Recognition, volume 2016-Decem. 2016;\hspace{0pt}
  1874--1883.

\bibitem{Zhang2018d}
Zhang Y, Li K, Li K, Wang L, Zhong B, Fu Y.
\newblock {Image Super-Resolution Using Very Deep Residual Channel Attention
  Networks}.
\newblock In Lecture Notes in Computer Science (including subseries Lecture
  Notes in Artificial Intelligence and Lecture Notes in Bioinformatics), volume
  11211 LNCS. 2018;\hspace{0pt} 294--310.

\bibitem{Haris2018}
Haris M, Shakhnarovich G, Ukita N.
\newblock {Deep Back-Projection Networks for Super-Resolution}.
\newblock In Proceedings of the IEEE Conference on Computer Vision and Pattern
  Recognition. 2018;\hspace{0pt} 1664--1673.

\bibitem{Shi2016}
Shi W, Caballero J, Theis L, Huszar F, Aitken A, Ledig C, Wang Z.
\newblock {Is the deconvolution layer the same as a convolutional layer?}
\newblock arXiv preprint arXiv:160907009 2016;\hspace{0pt}.

\bibitem{Trabelsi2017}
Trabelsi C, Bilaniuk O, Zhang Y, Serdyuk D, Subramanian S, Santos JF, Mehri S,
  Rostamzadeh N, Bengio Y, Pal CJ.
\newblock {Deep Complex Networks}.
\newblock arXiv preprint arXiv:170509792 2017;\hspace{0pt}.

\bibitem{Wang2018b}
Wang X, Yu K, Wu S, Gu J, Liu Y, Dong C, Qiao Y, Change~Loy C.
\newblock {ESRGAN: Enhanced Super-Resolution Generative Adversarial Networks}.
\newblock In Proceedings of the European Conference on Computer Vision
  Workshops. 2018;\hspace{0pt} .

\bibitem{Huang2017}
Huang G, Liu Z, Van Der~Maaten L, Weinberger KQ.
\newblock {Densely Connected Convolutional Networks}.
\newblock In Proceedings of the IEEE Conference on Computer Vision and Pattern
  Recognition. 2017;\hspace{0pt} 2261--2269.

\bibitem{Zhang2018c}
Zhang Y, Tian Y, Kong Y, Zhong B, Fu Y.
\newblock {Residual Dense Network for Image Super-Resolution}.
\newblock In Proceedings of the IEEE Computer Society Conference on Computer
  Vision and Pattern Recognition. 2018;\hspace{0pt} 2472--2481.

\bibitem{Mao2017}
Mao X, Li Q, Xie H, Lau RYK, Wang Z, Smolley SP.
\newblock {Least Squares Generative Adversarial Networks}.
\newblock In Proceedings of the IEEE International Conference on Computer
  Vision. 2017;\hspace{0pt} 2794--2802.

\bibitem{Ledig2017}
Ledig C, Theis L, Husz{\'{a}}r F, Caballero J, Cunningham A, Acosta A, Aitken
  A, Tejani A, Totz J, Wang Z, Shi~Twitter W.
\newblock {Photo-Realistic Single Image Super-Resolution Using a Generative
  Adversarial Network}.
\newblock In IEEE Conference on Computer Vision and Pattern Recognition.
  2017;\hspace{0pt} 4681--4690.

\bibitem{Ulyanov2018}
Ulyanov D, Vedaldi A, Lempitsky V.
\newblock {Deep Image Prior}.
\newblock In Proceedings of the IEEE Conference on Computer Vision and Pattern
  Recognition. 2018;\hspace{0pt} 9446--9454.

\bibitem{Mataev2019}
Mataev G, Elad M, Milanfar P.
\newblock {DeepRED: Deep Image Prior Powered by RED}.
\newblock arXiv preprint arXiv:190310176 2019;\hspace{0pt}.

\bibitem{Goodfellow2016}
Goodfellow I, Bengio Y, Courville A.
\newblock {Deep Learning}.
\newblock MIT Press, 2016.

\bibitem{Krizhevsky2012}
Krizhevsky A, Sutskever I, Geoffrey~E H.
\newblock {ImageNet Classification with Deep Convolutional Neural Networks}.
\newblock In Advances in Neural Information Processing Systems.
  2012;\hspace{0pt} 1097--1105.

\bibitem{Rother2004}
Rother C, Kolmogorov V, Blake A.
\newblock {GrabCut - Interactive Foreground Extraction Using Iterated Graph
  Cuts}.
\newblock In ACM SIGGRAPH, volume~23. 2004;\hspace{0pt} 309--314.

\bibitem{Friedman1937}
Friedman M.
\newblock {The Use of Ranks to Avoid the Assumption of Normality Implicit in
  the Analysis of Variance}.
\newblock Journal of the American Statistical Association
  1937;\hspace{0pt}32(200):675--701.

\bibitem{Wilcoxon1992}
Wilcoxon F.
\newblock {Individual Comparisons by Ranking Methods}.
\newblock In Biometrics Bulletin, volume~1. Springer, 1945;\hspace{0pt} 80.

\bibitem{Stikov2019}
Stikov N, Trzasko JD, Bernstein MA.
\newblock {Reproducibility and the future of MRI research}.
\newblock Magnetic Resonance in Medicine 2019;\hspace{0pt}82(6):1981--1983.

\bibitem{Uecker2015a}
Uecker M, Ong F, Tamir JI, Bahri D, Virtue P, Cheng JY, Zhang T, Lustig M.
\newblock {Berkeley Advanced Reconstruction Toolbox}.
\newblock In Proceedings of the International Society of Magnetic Resonance in
  Medicine. 2015;\hspace{0pt} 2486.

\bibitem{Shitrit2017}
Shitrit O, Riklin~Raviv T.
\newblock {Accelerated magnetic resonance imaging by adversarial neural
  network}.
\newblock In Lecture Notes in Computer Science (including subseries Lecture
  Notes in Artificial Intelligence and Lecture Notes in Bioinformatics), volume
  10553 LNCS. Springer, Cham, 2017;\hspace{0pt} 30--38.

\bibitem{Hammernik2018a}
Hammernik K, Kobler E, Pock T, Recht M, Sodickson DK, Knoll F.
\newblock {Variational Adversarial Networks for Accelerated MR Image
  Reconstruction}.
\newblock In Proceedings of the International Society of Magnetic Resonance in
  Medicine. 2018;\hspace{0pt} 1091.

\bibitem{Caballero2014a}
Caballero J, Bai W, Price AN, Rueckert D, Hajnal JV.
\newblock {Application-driven MRI: Joint Reconstruction and Segmentation From
  Undersampled MRI Data}.
\newblock In International Conference on Medical Image Computing and
  Computer-Assisted Intervention. Springer, 2014;\hspace{0pt} 106--113.

\bibitem{Schlemper2018}
Schlemper J, Oktay O, Bai W, Castro DC, Duan J, Qin C, Hajnal JV, Rueckert D.
\newblock {Cardiac MR Segmentation From Undersampled k-Space Using Deep Latent
  Representation Learning}.
\newblock In International Conference on Medical Image Computing and
  Computer-Assisted Intervention. Springer, 2018;\hspace{0pt} 259--267.

\bibitem{Antun2019}
Antun V, Renna F, Poon C, Adcock B, Hansen AC.
\newblock {On instabilities of deep learning in image reconstruction-Does AI
  come at a cost?}
\newblock arXiv preprint arXiv:190205300 2019;\hspace{0pt}.

\end{thebibliography}

\ifbDraft
\clearpage \section{Supplementary Material}

\sfigI
\sfigII
\sfigIII
\losf

\stabI
\lost
\else
\figI
\figII
\tabI
\figIII
\figIV
\figV
\tabII
\figVI

\sfigI
\sfigII
\sfigIII

\stabI
\clearpage
\section{List of Figures}
\lof
%
\section{List of Tables}
\lot
%
\section{List of Supporting Figures}
\losf
%

\section{List of Supporting Tables}
\lost


\setboolean{bDraft}{true}
\setcounter{figure}{0}
\setcounter{sfigure}{0}
\setcounter{table}{0}
\setcounter{stable}{0}

\figI
\figII
\tabI
\figIII
\figIV
\figV
\tabII
\figVI

\clearpage
\sfigI
\sfigII
\sfigIII

\stabI
\fi

\end{document}